\documentclass[aps,pra,twocolumn,showpacs,notitlepage,superscriptaddress,letterpaper]{revtex4-1}

\usepackage{graphicx, amsmath, amssymb, amsfonts, pifont, bm, cancel}
\usepackage[usenames]{color}
\usepackage{subfigure}
\usepackage[normalem]{ulem} 
\usepackage{circuitikz}
\usepackage{hyperref}
\usepackage{threeparttable}
\usepackage{array}
\newcolumntype{P}[1]{>{\centering\arraybackslash}p{#1}}
\newcolumntype{M}[1]{>{\centering\arraybackslash}m{#1}}

\definecolor{MyDarkBlue}{rgb}{0,0,1}

\newcommand{\beq}{\begin{equation}}
\newcommand{\eeq}{\end{equation}}
\newcommand{\bel}{\begin{align*}}
\newcommand{\tamam}{\end{align*}}

\newcommand{\ket}[1]{|#1\rangle}
\newcommand{\brak}[1]{\langle#1|}

\newcommand{\beqa}{\begin{eqnarray}}             
\newcommand{\eeqa}{\end{eqnarray}}               

\newcommand{\ain}{\hat{a}_\text{in}}

\newcommand{\aout}{\hat{a}_\text{out}}

\newcommand{\eprobe}{|\text{e}\rangle_{\text{p}}}
\newcommand{\gprobe}{|\text{g}\rangle_{\text{p}}}
\newcommand{\Emirror}{|\text{E}\rangle}
\newcommand{\Dmirror}{|\text{D}\rangle}

\newcommand{\Gmirror}{|\text{G}\rangle}

\newcommand{\Pp}{P_\text{p}} 
\newcommand{\omegap}{\omega_\text{p}}
\newcommand{\omegaq}{\omega_\text{q}}
\newcommand{\Omegap}{\Omega_\text{p}} 
\newcommand{\domega}{\delta\omega}
\newcommand{\nth}{\bar{n}_\text{th}}
\newcommand{\kB}{k_\text{B}} 

\newcommand{\sigmam}{\hat{\sigma}_-}
\newcommand{\sigmap}{\hat{\sigma}_+}

\newcommand{\sigmaz}{\hat{\sigma}_z}

\newcommand{\drhoee}{\dot{\rho}_\text{e,e}}
\newcommand{\drhoeg}{\dot{\rho}_\text{e,g}}
\newcommand{\drhoge}{\dot{\rho}_\text{g,e}}
\newcommand{\drhogg}{\dot{\rho}_\text{g,g}}

\newcommand{\rhoee}{\rho_\text{e,e}}
\newcommand{\rhoeg}{\rho_\text{e,g}}
\newcommand{\rhoge}{\rho_\text{g,e}}
\newcommand{\rhogg}{\rho_\text{g,g}}
\newcommand{\rhoeess}{\rho_\text{e,e}^\text{ss}}
\newcommand{\rhoegss}{\rho_\text{e,g}^\text{ss}}

\newcommand{\drhosub}[1]{\dot{\rho}_\text{#1}}
\newcommand{\rhosub}[1]{\rho_\text{#1}}

\newcommand{\Gammaloss}{\Gamma_\text{loss}} 
\newcommand{\Gammaprime}{\Gamma^\prime} 
\newcommand{\Gammaphi}{\Gamma_\varphi} 
\newcommand{\Gammaoneth}{\Gamma_{1}^\text{th}}
\newcommand{\Gammatwoth}{\Gamma_{2}^\text{th}}
\newcommand{\GammaoneD}{\Gamma_\text{1D}}
\newcommand{\GammaoneDp}{\Gamma_\text{1D,p}}
\newcommand{\Gammaone}{\Gamma_\text{1}}
\newcommand{\Gammatwo}{\Gamma_\text{2}}

\newcommand{\sub}[2]{{#1}_\text{#2}}

\begin{document}

\title{Waveguide-mediated interaction of artificial atoms in the strong coupling regime} 

\author{Mohammad~Mirhosseini}
\thanks{These authors contributed equally to this work.}
\author{Eunjong~Kim}
\thanks{These authors contributed equally to this work.}
\author{Xueyue~Zhang}
\author{Alp~Sipahigil}
\author{Paul~B.~Dieterle}
\author{Andrew~J.~Keller}
\affiliation{Kavli Nanoscience Institute and Thomas J. Watson, Sr., Laboratory of Applied Physics, California Institute of Technology, Pasadena, California 91125, USA.}
\affiliation{Institute for Quantum Information and Matter, California Institute of Technology, Pasadena, California 91125, USA.}
\author{Ana~Asenjo-Garcia}
\affiliation{Institute for Quantum Information and Matter, California Institute of Technology, Pasadena, California 91125, USA.}
\affiliation{Norman Bridge Laboratory of Physics, California Institute of Technology, Pasadena, California 91125, USA.}
\affiliation{ICFO-Institut de Ciencies Fotoniques, The Barcelona Institute of Science and Technology, 08860 Castelldefels (Barcelona), Spain}
\author{Darrick~E.~Chang}
\affiliation{ICFO-Institut de Ciencies Fotoniques, The Barcelona Institute of Science and Technology, 08860 Castelldefels (Barcelona), Spain}
\affiliation{ICREA-Instituci\'{o} Catalana de Recerca i Estudis Avan\c{c}ats, 08015 Barcelona, Spain}
\author{Oskar~Painter}
\email{opainter@caltech.edu}
\homepage{http://copilot.caltech.edu}
\affiliation{Kavli Nanoscience Institute and Thomas J. Watson, Sr., Laboratory of Applied Physics, California Institute of Technology, Pasadena, California 91125, USA.}
\affiliation{Institute for Quantum Information and Matter, California Institute of Technology, Pasadena, California 91125, USA.}

\date{\today}


\begin{abstract} 
{Waveguide quantum electrodynamics studies photon-mediated interactions of quantum emitters in a one-dimensional radiation channel. Although signatures of such interactions have been observed previously in a variety of physical systems, observation of coherent cooperative dynamics has been obscured by radiative decay of atoms into the waveguide. Employing transmon qubits as artificial atoms coupled to a microwave coplanar waveguide, here we observe dynamical oscillations in an open system where a designated probe qubit interacts with an entangled dark state of an array of qubits which effectively traps radiation as an atomic cavity. The qubit-cavity system is shown to achieve a large cooperativity of $\mathcal{C}=172$ due to collective enhancement of photon-mediated interactions, entering the strong coupling regime.  The quantum coherence of the dark state cavity is also explored through its nonlinear response at the single-excitation level.  With realistic refinements, this system is suitable for studying the many-body dynamics of large ($N>10$) quantum spin chains, synthesizing highly non-classical radiation fields on demand, and implementing universal quantum logic operations with high fidelity on information encoded within decoherence-free subspaces.}
\end{abstract}
\maketitle
                                                                                                                   
Cooperative interaction of atoms in the presence of a radiation field has been studied since the early days of quantum physics. The hallmark of such effects is the formation of super- and sub-radiant states in the spontaneous emission of an ensemble of atoms, first studied by Dicke~\cite{Dicke:1954bl}. While super-radiance deals with the exchange of real photons between atoms within an electromagnetic reservoir~\cite{Gross:1982js,Eberly:2006jh,Scully:2007cw}, cooperative effects can also be achieved by exchange of virtual photons. A virtual photon emitted by an atom can be re-absorbed by another identical atom, giving rise to an effective exchange-type interaction between a pair of resonant atoms~\cite{Osnaghi:2001bw,Majer:2007ema,Zheng:2000eo,Friedberg:2008ch}. The presence of photon-mediated interactions can be identified as an energy level shift of a collective excited state of an atomic ensemble, a phenomenon known as the cooperative Lamb shift~\cite{Rohlsberger:2010jq,Keaveney:2012gm,Meir:2014fw}. However, dynamical signatures of such interactions are difficult to observe in an open system with optically small ensembles (atomic separation much smaller than radiation wavelength, $d\ll \lambda_{0}$) because the evolution of the system is dominated by cooperative decay and short-ranged dipole interactions~\cite{Friedberg:1973hr,Scully:2009kq}.

Progress in the preparation and control of individual quantum emitters has recently revived interest in studying cooperative effects in waveguide quantum electrodynamics (QED) systems~\cite{Roy:2017hn}. In this physical system, atoms interact via a one-dimensional (1D) radiation channel supporting a continuum of electromagnetic modes, such as an optical fiber or a microwave waveguide~\cite{Lodahl:2015fy,Gu:2017et}. The confinement of the electromagnetic field in a waveguide can lead to full collection of radiative emission from the coupled atoms, which in the extreme case manifests as full extinction of transmitted light through the waveguide due to interference with atomic resonance fluorescence~\cite{Astafiev:2010cm}. Additionally, the waveguide-mediated interaction becomes long ranged in a 1D system~\cite{Dzsotjan:2010,Gonzalez-Tudela:2011,AsenjoGarcia:2017bm} which allows for studying situations where the emitters are separated over distances $d \gtrsim \lambda_{0}$. The spatial arrangement of emitters in such extensive states has been shown to lead to qualitatively different regimes. For instance, for a separation $d = \lambda_{0}/4$ of identical emitters the cooperative coherent interactions between emitters is maximized while cooperative decay is completely suppressed~\cite{vanLoo:2013df}. Single atom decay into the waveguide is still present, however, which obscures the waveguide-mediated interactions regardless of emitter separation~\cite{vanLoo:2013df,Kockum:2018hs}. A number of methods for circumventing this issue and protecting against radiative decay in an open waveguide system have been considered, including terminating the waveguide with a mirror~\cite{Eschner:2001ib,Hoi:2015fh}, loading the waveguide periodically to create photonic bandgaps~\cite{Liu:2016ic,Sundaresan:2018vf,Mirhosseini:2018wg}, using ``giant atoms" with multiple coupling points to a waveguide~\cite{Kockum:2018hs}, and encoding quantum information in a decoherence-free subspace of multiple atoms~\cite{Lidar:1998br,Paulisch:2016ib}.


\begin{figure*}[t!]
\begin{center}
\includegraphics[width=1\textwidth]{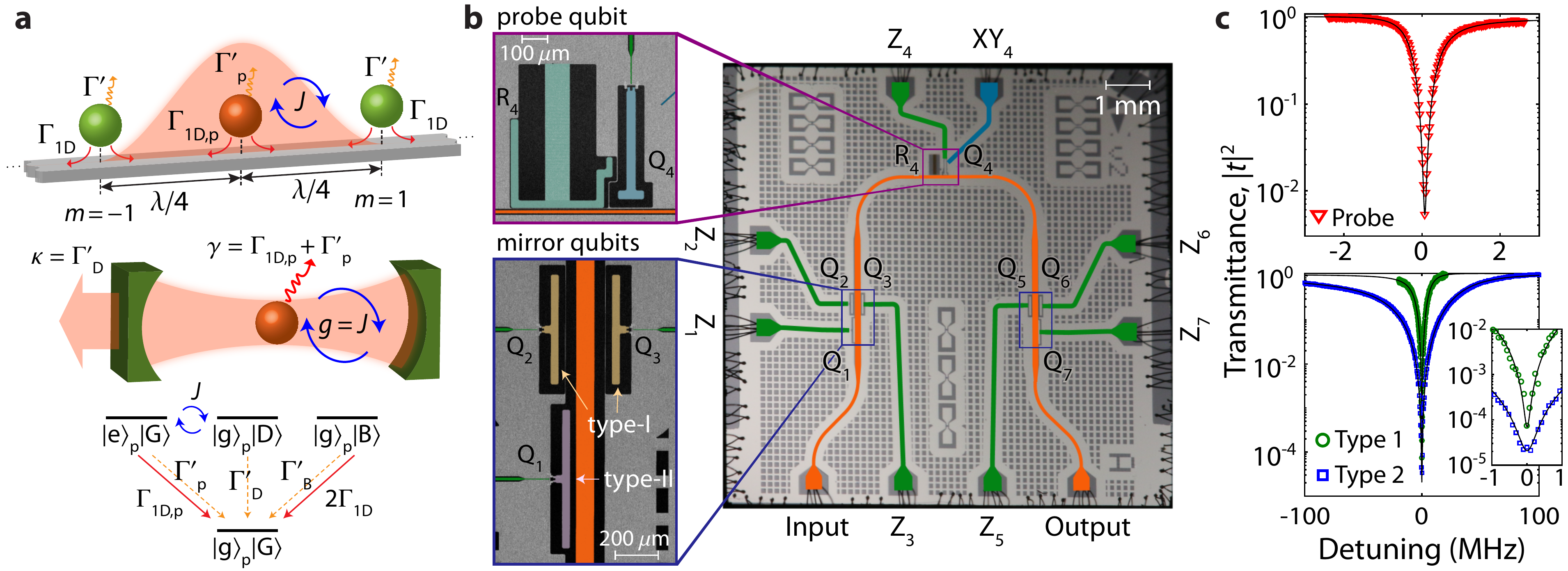}
\caption{\textbf{Waveguide QED setup.} \textbf{a}, Top: schematic showing cavity configuration of waveguide QED system consisting of an array of $N$ mirror qubits ($N=2$ shown; green) coupled to the waveguide with an inter-qubit separation of $\lambda_{0}/2$, with a probe qubit (red) at the center of the mirror array. Middle: schematic showing analogous cavity QED system with correspondence to waveguide parameters. Bottom: energy level diagram of the system of three qubits (2 mirror, one probe). The mirror dark state $\Dmirror$ is coupled to the excited state of the probe qubit $\eprobe$ at a cooperatively enhanced rate of $2J = \sqrt{2\Gamma_\text{1D}\Gamma_\text{1D,p}}$. The bright state is decoupled from the probe qubit. \textbf{b}, Optical image of the fabricated waveguide QED chip. Tunable transmon qubits interact via microwave photons in a superconducting coplanar waveguide (CPW; false-color orange trace). The CPW is used for externally exciting the system and is terminated in a $50$-$\Omega$ load. Insets: Scanning electron microscope image of the different qubit designs used in our experiment. The probe qubit, designed to have $\Gamma_\text{1D,p}/2\pi = 1$~MHz, is accessible via a separate CPW (XY$_4$; false-color blue trace) for state preparation, and is also coupled to a compact microwave resonator (R$_{4}$; false-color cyan) for dispersive readout. The mirror qubits come in two types: type-I with $\Gamma_\text{1D}/2\pi = 20$~MHz and type-II with $\Gamma_\text{1D}/2\pi = 100$~MHz. \textbf{c}, Waveguide transmission spectrum across individual qubit resonances (top: probe qubit (Q$_4$) ; bottom: individual type-I (Q$_6$, green curve) and type-II (Q$_1$, blue curve) mirror qubits). From Lorentzian lineshape fit of the measured waveguide transmission spectra we infer Purcell factors of $P_\text{1D} = 11$ for the probe qubit and $P_\text{1D} = 98$ ($219$) for the type-I (type-II) mirror qubit.}\label{fig:schematic} 
\end{center}
\end{figure*}

Here we explore the dynamical properties of an engineered superconducting quantum circuit consisting of an array of artificial atoms, in the form of transmon qubits~\cite{Koch2007}, which are strongly coupled to a common microwave waveguide channel.  Inspired by recent theoretical ideas~\cite{Chang:2012co,Albrecht:je}, we use precise control of the phase separation of the qubits along the waveguide and an ancillary probe qubit to create a collective sub-radiant state of the qubit array in which radiative decay is strongly suppressed while interactions with the probe qubit via microwave photons in the waveguide are cooperatively enhanced. Using a combination of waveguide transmission and individual addressing of the probe qubit we are able to observe spectroscopic and time-domain signatures of the collective dynamics of the qubit array. This yields direct evidence of strong coupling between the probe qubit and the sub-radiant state, while also providing a sensitive measure of the coherence and quantum nonlinear behavior of the collective states of the qubit array. 


The collective evolution of an array of resonant qubits coupled to a 1D waveguide can be formally described by a master equation of the form $\dot{\hat{\rho}} = -i/\hbar[\hat{H}_{\text{eff}},\hat{\rho}] + \sum_{m,n} \Gamma_{m,n} \hat{\sigma}_\text{ge}^m \hat{\rho} \hat{\sigma}_\text{eg}^n$~\cite{Chang:2012co,Lalumiere:2013io}, where $\hat{\sigma}_\text{ge}^m = \ket{\text{g}_{m}}\brak{\text{e}_{m}}$, and $m$ and $n$ represent indices into the qubit array. Within the Born-Markov approximation, the effective Hamiltonian can be written in the interaction picture as

\begin{align}
\hat{H}_{\text{eff}} = \hbar \sum_{m,n} \left(J_{m,n}-i\frac{\Gamma_{m,n}}{2}\right)\hat{\sigma}_\text{eg}^m \hat{\sigma}_\text{ge}^n,
\label{eq:Heff}
\end{align}

\noindent where $J_{m,n} = \Gamma_\text{1D} \sin{(k_{0}|x_m - x_n|)}/2$ and $\Gamma_{m,n} = \Gamma_\text{1D} \cos{(k_{0}|x_m - x_n|)}$ denote the cooperative exchange interaction and cooperative dissipation terms, respectively.  $x_{m(n)}$ are the spatial locations of the qubits along the waveguide propagation axis, $\Gamma_\text{1D}$ is the energy decay rate of individual qubits into the waveguide, and $k_{0} = \omega_{0}/c$ is the wavenumber of the guided mode of the waveguide at resonance ($\omega_{0}$) with the qubits. The collective states of the coupled qubits, along with their corresponding frequency shifts and decay rates, can be found by diagonalizing the effective Hamiltonian. 

Figure~\ref{fig:schematic}(a) depicts the waveguide QED system considered in this work.  The system consists of an array of $N$ qubits separated by distance $d = \lambda_{0}/2$ and a separate probe qubit centered in the middle of the array with waveguide decay rate $\Gamma_\text{1D,p}$. In this configuration, the effective Hamiltonian can be simplified in the single-excitation manifold to

\begin{multline}
\hat{H}_{\text{eff}} =  -\frac{iN\hbar\Gamma_\text{1D}}{2} \hat{S}_\text{B}^\dagger\hat{S}_\text{B} -\frac{i\hbar\Gamma_\text{1D,p}}{2} \hat{\sigma}^\text{(p)}_\text{ee} \\
+ \hbar J\left({\hat{\sigma}_\text{ge}^\text{(p)}}\hat{S}_\text{D}^\dagger  + h. c.\right),
\label{eq:Heffdiag}
\end{multline}

\noindent where $\hat{S}_\text{B,D} = 1/\sqrt{N} \sum_{m>0} (\hat{\sigma}_\text{ge}^{m} \mp \hat{\sigma}_\text{ge}^{-m}){(-1)}^m $ are the lowering operators of the bright collective state and  the fully-symmetric dark collective state of the qubit array (as shown in Fig.~\ref{fig:schematic}(a), $m>0$ and $m<0$ denote qubits to the right and left of the probe qubit, respectively). As evident by the last term in the Hamiltonian, the probe qubit is coupled to this dark state at a cooperatively enhanced rate $2J = \sqrt{N}\sqrt{\Gamma_\text{1D}\Gamma_\text{1D,p}}$. The bright state super-radiantly emits into the waveguide at a rate of $N\Gamma_{\text{1D}}$. The collective dark state has no coupling to the waveguide, and a decoherence rate $\Gamma^{\prime}_{\text{D}}$ which is set by parasitic damping and dephasing not captured in the simple waveguide QED model (see App.~\ref{App:C}). In addition to the bright and dark collective states described above, there exist an additional $N-2$ collective states of the qubit array with no coupling to either the probe qubit or the waveguide~\cite{Chang:2012co}.

The subsystem consisting of coupled probe qubit and symmetric dark state of the mirror qubit array can be described in analogy to a cavity QED system~\cite{Chang:2012co}. In this picture the probe qubit plays the role of a two-level atom and the dark state mimics a high-finesse cavity with the qubits in the $\lambda_{0}/2$-spaced array acting as atomic mirrors (see Fig.~\ref{fig:schematic}(a)). In general, provided that the fraction of excited array qubits remains small as $N$ increases, $\hat{S}_\text{D}$ stays nearly bosonic and the analogy to the Jaynes-Cummings model remains valid. Mapping waveguide parameters to those of a cavity QED system, the cooperativity between probe qubit and atomic cavity can be written as $\mathcal{C} = (2J)^2/(\Gamma_\text{1D,p} + \Gamma^{\prime}_{\text{p}}) \Gamma^{\prime}_{\text{D}} \approx N P_\text{1D}$. Here $P_\text{1D} = \Gamma_\text{1D}/\Gamma^{\prime}$ is the single qubit Purcell factor, which quantifies the ratio of waveguide emission rate to parasitic damping and dephasing rates. Attaining $\mathcal{C}>1$ is a prerequisite for observing coherent quantum effects.  Referring to the energy level diagram of Fig.~\ref{fig:schematic}(a), by sufficiently reducing the waveguide coupling rate of the probe qubit one can also realize a situation in which $J>(\Gamma_\text{1D,p} + \Gamma^{\prime}_{\text{p}}),\Gamma^{\prime}_{\text{D}}$, corresponding to the strong coupling regime of cavity QED between excited state of the probe qubit ($\eprobe\Gmirror$) and a single photon in the atomic cavity ($\gprobe\Dmirror$).

The fabricated superconducting circuit used to realize the waveguide QED system is shown in Fig.~\ref{fig:schematic}(b). The circuit consists of seven transmon qubits (Q$_j$ for $j = 1$-$7$), all of which are side-coupled to the same coplanar waveguide (CPW).  Each qubit's transition frequency is tunable via an external flux bias port (Z$_1$-Z$_7$). We use the top-center qubit in the circuit (Q$_4$) as a probe qubit.  This qubit can be independently excited via a weakly-coupled CPW drive line (XY$_4$), and is coupled to a lumped-element microwave cavity (R$_4$) for dispersive readout of its state. The other six qubits are mirror qubits.  The mirror qubits come in two different types (I and II), which have been designed to have different waveguide coupling rates ($\Gamma_{\text{1D,I}}/2\pi = 20$~MHz and $\Gamma_{\text{1D,II}}/2\pi = 100$~MHz) in order to provide access to a range of Purcell factors. Type-I mirror qubits also lie in pairs across the CPW waveguide and have rather large ($\sim 50$~MHz) direct coupling.  We characterize the waveguide and parasitic coupling rates of each individual qubit by measuring the phase and amplitude of microwave transmission through the waveguide (see Fig.~\ref{fig:schematic}(c)).  In order to reduce thermal noise, measurements are performed in a dilution refrigerator at a base temperature of $8$~mK (see App.~\ref{App:A} for set-up).  For a sufficiently weak coherent drive the effects of qubit saturation can be neglected and the on-resonance extinction of the coherent waveguide tone relates to a lower bound on the individual qubit Purcell factor. Any residual waveguide thermal photons~\cite{Yan:2017fp}, however, can result in weak saturation of the qubit and a reduction of the on-resonance extinction.  We find an on-resonance intensity transmittance as low as $2\times10^{-5}$ for the type-II mirror qubits, corresponding to an upper bound on the CPW mode temperature of $43$~mK and a lower bound on the Purcell factor of $200$.  Further details of the design, fabrication, and measured parameters of probe and each mirror qubit are provided in App.~\ref{App:A} and App.~\ref{App:B}.

\begin{figure}[t]
\begin{center}
\includegraphics[width=\columnwidth]{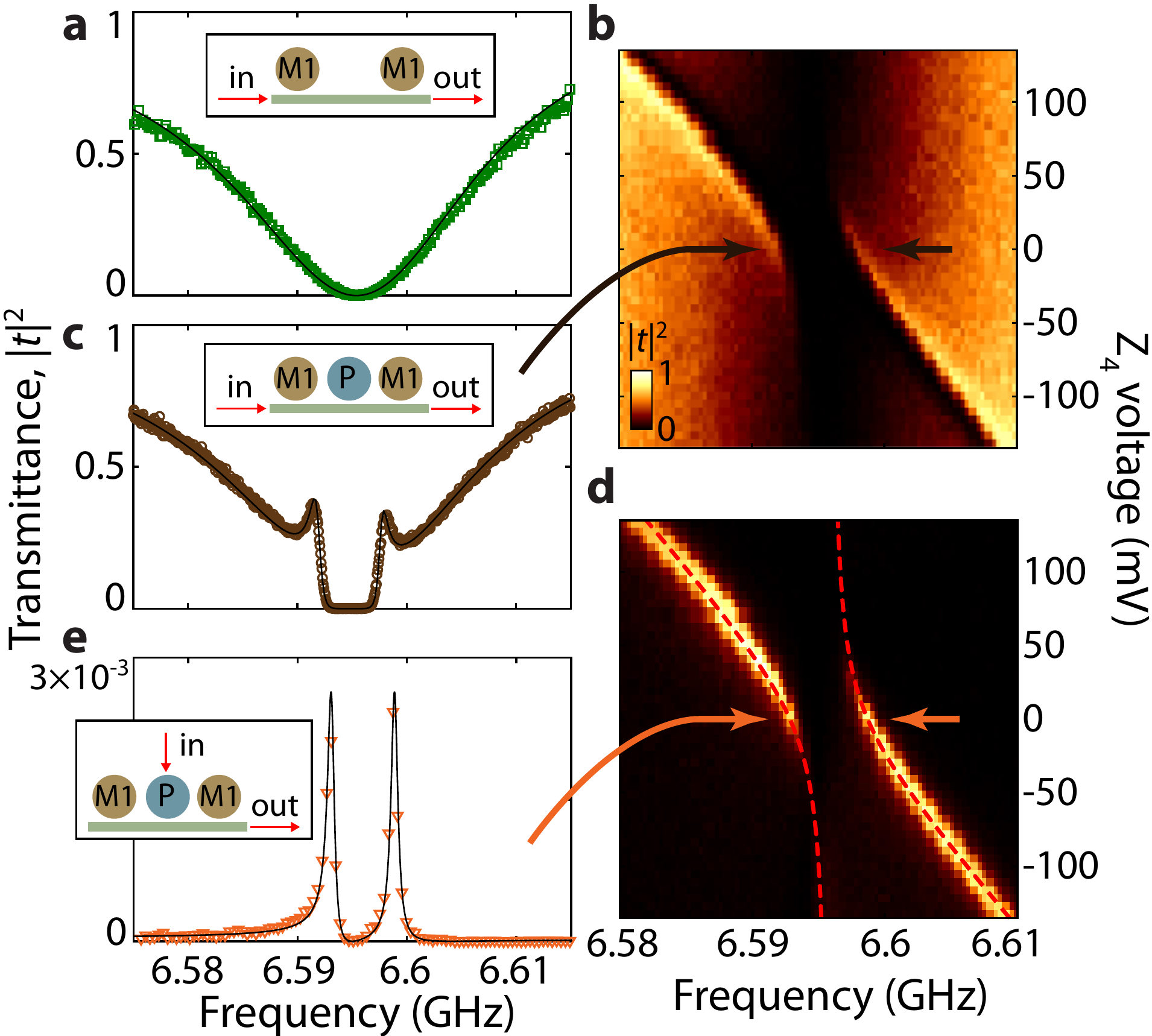}
\caption{\textbf{Vacuum Rabi splitting.} \textbf{a}, Transmission through the waveguide for two mirror qubits (Q$_2$,Q$_6$) on resonance, with the remaining qubits on the chip tuned away from the measurement frequency range. \textbf{b}, Transmission through the waveguide as a function of the flux bias tuning voltage of the probe qubit (Q$_4$). \textbf{c}, Waveguide transmission spectrum for the three qubits tuned into resonance. \textbf{d}. Transmission spectrum as measured between the probe qubit drive line XY$_4$ and the waveguide output as a function of flux bias tuning of the probe qubit. \textbf{e}, XY$_4$-to-waveguide transmission spectrum for the three qubits tuned into resonance. The dashed red lines in (d) and solid black line in (e) show predictions of a numerical model with experimentally measured qubit parameters. The prediction in (e) includes slight power broadening effects.  Legend: M1 and P denote type-I mirror qubits (Q$_2$,Q$_6$) and the probe qubit (Q$_4$), respectively.}
\label{fig:spectrum} 
\end{center}
\end{figure}

 \begin{figure*}[t]
\begin{center}
\includegraphics[width=\textwidth]{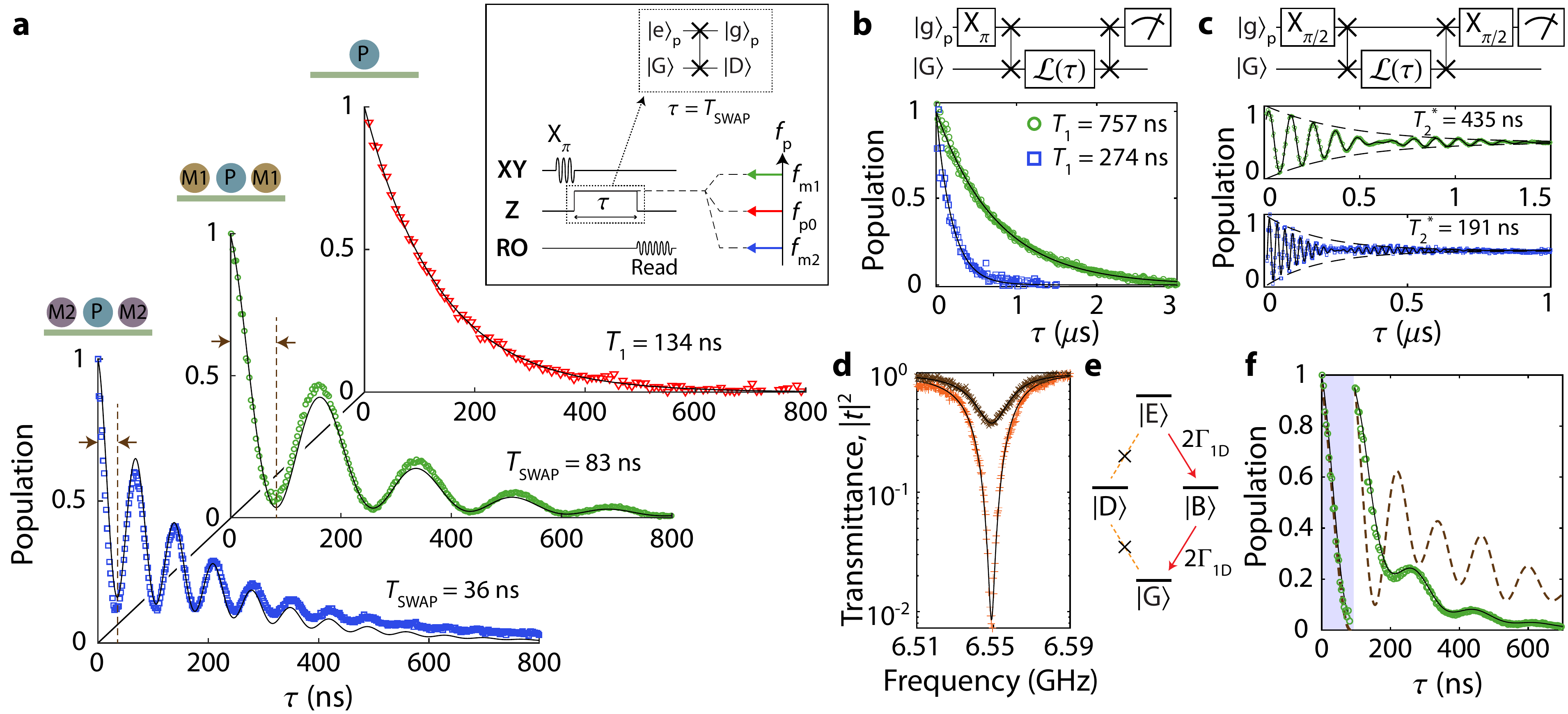}
\caption{\textbf{Vacuum Rabi oscillations.}  \textbf{a}, Measured population of the excited state of the probe qubit for three different scenarios.  (i) Probe qubit tuned to $f_{\text{p0}} = 6.55$~GHz, with all mirror qubits tuned away, corresponding to free population decay (red curve). (ii) Probe qubit tuned into resonance with a pair of type-I mirror qubits (Q$_2$,Q$_6$) at frequency $f_{\text{m1}} = 6.6$~GHz corresponding to $d_{\text{I}} = \lambda_{0}/2$ (green curve). (iii) Probe qubit tuned in resonance with type-II mirror qubits (Q$_1$,Q$_7$) at frequency $f_{\text{m2}} = 5.826$~GHz corresponding to $d_{\text{II}} = \lambda_{0}/2$ (blue curve).  Inset: The sequence of pulses applied during the measurement. Legends: P, M1, and M2 denote the probe qubit, type-I, and type-II mirror qubits, respectively. \textbf{b}, Measurement of the population decay time ($T_{1,\text{D}}$) of the dark state of type-I (green curve) and type-II (blue curve) mirror qubits. \textbf{c}, Corresponding Ramsey coherence time ($T_{2,\text{D}}^*$) of the type-I and type-II dark states. \textbf{d}, Waveguide transmission spectrum through the atomic cavity without (red data points) and with (orange data points) pre-population of the cavity. Here the atomic cavity was initialized in a single photon state by performing an iSWAP gate with the probe qubit followed by detuning of the probe qubit away from resonance. In both cases the transmission measurement is performed using coherent rectangular pulses with a duration of $260$~ns and a peak power of $P \approx 0.03 (\hbar \omega_{0} \Gamma_\text{1D})$. Solid lines show theory fits from numerical modeling of the system. \textbf{e},  Energy level diagram of the 0 ($\ket{\text{G}}$), 1 ($\ket{\text{D}}$,$\ket{\text{B}}$), and 2 ($\ket{\text{E}}$) excitation manifolds of the atomic cavity indicating waveguide induced decay and excitation pathways.  \textbf{f}, Rabi oscillation with two excitations in the system of probe qubit and atomic cavity. The shaded region shows the first iSWAP step in which an initial probe qubit excitation is transferred to the atomic cavity. Populating the probe qubit with an additional excitation at this point results in strong damping of subsequent Rabi oscillations due to the rapid decay of state $\ket{\text{E}}$. Dashed brown curve is the predicted result for interaction of the probe qubit with an equivalent linear cavity. In (d)-(f) the atomic cavity is formed from type-I mirror qubits Q$_2$ and Q$_6$.}\label{fig:Rabi} 
\end{center}
\end{figure*}

 The transmission through the waveguide, in the presence of the probe qubit, can also be used to measure spectroscopic signatures of the collective dark state of the qubit array. As an example of this we utilize a single pair of mirror qubits (Q$_2$, Q$_6$ of type-I), which we tune to a frequency where their separation along the waveguide axis is $d = \lambda_{0}/2$. The remaining qubits on the chip are decoupled from the waveguide input by tuning their frequency away from the measurement point.  Figure~\ref{fig:spectrum}(a) shows the waveguide transmission spectrum for a weak coherent tone in which a broad resonance dip is evident corresponding to the bright state of the mirror qubit pair. We find a bright state waveguide coupling rate of $\Gamma_\text{1D,B}\approx 2\Gamma_\text{1D} = 2\pi \times 26.8$~MHz by fitting a Lorentizan lineshape to the spectrum. The dark state of the mirror qubits, being dark, is not observable in this waveguide spectrum. The dark state becomes observable, however, when measuring the waveguide transmission with the probe qubit tuned into resonance with the mirror qubits (see Fig.~\ref{fig:spectrum}(b)). In addition to the broad response from the bright state, in this case there appears two spectral peaks near the center of the bright state resonance (Fig.~\ref{fig:spectrum}(c)).  This pair of highly non-Lorentzian spectral features result from the Fano interference~\cite{Fano:1961ha} between the broad bright state and the hybridized polariton resonances formed between the dark state of the mirror qubits (atomic cavity photon) and the probe qubit. The hybridized probe qubit and atomic cavity eigenstates can be more clearly observed by measuring the transmission between the probe qubit drive line (XY$_4$) and the output port of the waveguide (see Fig.~\ref{fig:spectrum}(d)). As the XY$_4$ line does not couple to the bright state due to the symmetry of its positioning along the waveguide, we observe two well-resolved resonances in Fig.~\ref{fig:spectrum}(e) with mode splitting $2J/2\pi\approx 6$~MHz when the probe qubit is nearly resonant with the dark state.  Observation of vacuum Rabi splitting in the hybridized atomic cavity-probe qubit polariton spectrum signifies operation in the strong coupling regime.

 To further investigate the signatures of strong coupling we perform time domain measurements in which we prepare the system in the initial state $\gprobe\Gmirror \rightarrow \eprobe\Gmirror$ using a $10$~ns microwave $\pi$ pulse applied at the XY$_4$ drive line.  Following excitation of the probe qubit we use a fast flux bias pulse to tune the probe qubit into resonance with the collective dark state of the mirror qubits (atomic cavity) for a desired interaction time, $\tau$.  Upon returning to its initial frequency after the flux bias pulse, the excited state population of the probe qubit state is measured via the dispersively coupled readout resonator.  In Fig.~\ref{fig:Rabi}(a) we show a timing diagram and plot three measured curves of the probe qubit's population dynamics versus $\tau$.  The top red curve corresponds to the measured probe qubit's free decay, where the probe qubit is shifted to a detuned frequency $f_{\text{p0}}$ to eliminate mirror qubit interactions. From an exponential fit to the decay curve we find a decay rate of $1/T_1 \approx 2\pi \times 1.19$ MHz, in agreement with the result from waveguide spectroscopy at $f_{\text{p0}}$.  In the middle green and bottom blue curves we plot the measured probe qubit population dynamics when interacting with an atomic cavity formed from type-I and type-II mirror qubit pairs, respectively. In both cases the initially prepared state $\eprobe\Gmirror$ undergoes vacuum Rabi oscillations with the dark state of the mirror qubits $\gprobe\Dmirror$.  Along with the measured data we plot a theoretical model where the waveguide coupling, parasitic damping, and dephasing rate parameters of the probe qubit and dark state are taken from independent measurements, and the detuning between probe qubit and dark state is left as a free parameter (see App.~\ref{App:C}).  From the excellent agreement between measurement and model we infer an interaction rate of $2J/2\pi = 5.64$~MHz ($13.0$~MHz) and a cooperativity of $\mathcal{C} = 94$ ($172$) for the type-I (type-II) mirror system. For both mirror types we find that the system is well into the strong coupling regime ($J \gg \Gamma_{\text{1D,p}} + \Gamma^{\prime}_{\text{p}},\Gamma^{\prime}_{\text{D}}$), with the photon-mediated interactions dominating the decay and dephasing rates by roughly two orders of magnitude.

\begin{figure}[t!]
\begin{center}
\includegraphics[width=\columnwidth]{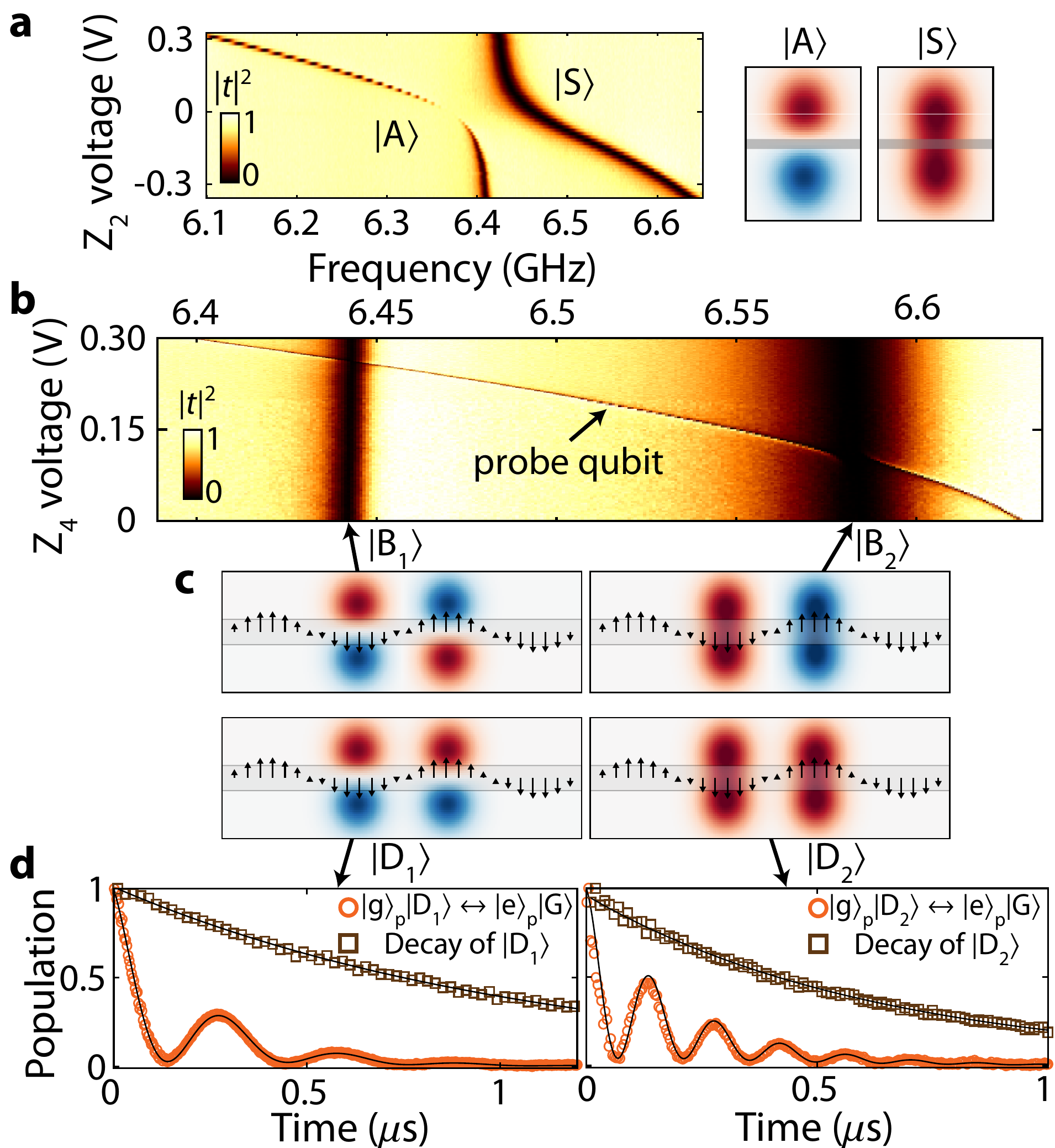}
\caption{\textbf{$N=4$ compound atomic mirrors.} \textbf{a}, Avoided mode crossing of a pair of type-I mirror qubits positioned on the opposite sides of the CPW.  Near the degeneracy point, the qubits form a pair of compound eigenstates consisting of symmetric ($\ket{\text{S}}$) and anti-symmetric ($\ket{\text{A}}$) states with respect to the waveguide axis.  \textbf{b}, Measured transmission through the waveguide with the pair of compound atomic mirrors aligned in frequency. The two broad resonances correspond to super-radiant states $\ket{\text{B}_{1}}$ and $\ket{\text{B}_{2}}$ as indicated. Tuning the probe qubit we observe signatures of the interaction of the probe qubit with the dark states. \textbf{c}, Illustration of the single-excitation manifold of the collective states of a $N=4$ mirror qubits forming a pair of compound atomic cavities. The bright (super-radiant) and dark (sub-radiant) states can be identified by comparing the symmetry of the compound qubit states with the resonant radiation field pattern in the waveguide.  \textbf{d}, Probe qubit measurements of the two dark states, $\ket{\text{D}_{1}}$ and $\ket{\text{D}_{2}}$.  In these measurements the frequency of each dark state is shifted to ensure $\lambda_{0}/2$ separation between the two compound atomic mirrors.}\label{fig:Compound} 
\end{center}
\end{figure}

The tunable interaction time in our measurement sequence also allows for performing state transfer between the probe qubit and the dark state of the mirror qubits via an iSWAP gate~\cite{Nielsen:2010vn}. We measure the dark state's population decay in a protocol where we excite the probe qubit and transfer the excitation into the dark state (see Fig.~\ref{fig:Rabi}(b)). From an exponential fit to the data we find a dark state decay rate of $T_{1,\text{D}} = 757$~ns ($274$~ns) for type-I (type-II) mirror qubits, enhanced by roughly the cooperativity over the bright state lifetime. In addition to lifetime, we can measure the coherence time of the dark state with a Ramsey-like sequence (see Fig.~\ref{fig:Rabi}(c)), yielding $T_{2,\text{D}}^* = 435$~ns ($191$~ns) for type-I (type-II) mirror qubits. The collective dark state coherence time, being slightly shorter than its population decay time, hints at correlated sources of noise in the distantly entangled qubits forming the dark state (see discussion in App.~\ref{App:D}). 


Our experiments so far have probed the atomic cavity system in the single excitation regime. Of course the atomic mirrors forming the cavity are not regular mirrors, and the two-level nature of the constituent qubit components leads to a number of interesting phenomena when the system is probed beyond the single-excitation manifold. As an example of the highly nonlinear behavior of the atomic mirrors, we populate the atomic cavity with a single photon via an iSWAP gate and then measure the waveguide transmission of weak coherent pulses through the system. Figure\,\ref{fig:Rabi}(d) shows transmission through the atomic cavity formed from type-I mirror qubits before and after adding a single photon. The sharp change in the transmissivity of the atomic cavity is a result of trapping in the long-lived dark state of the mirror qubits. The dark state has no transition dipole to the waveguide channel (see Fig.~\ref{fig:Rabi}(e)), and thus it cannot participate in absorption or emission of photons when probed via the waveguide.  As a result, populating the atomic cavity with a single photon makes it nearly transparent to incoming waveguide signals for the duration of the dark-state lifetime. This is analogous to the electron shelving phenomenon which leads to suppression of resonance fluorescence in three-level atomic systems~\cite{Cook:1985ft,Pegg:1986ha}. As a further example, we use the probe qubit to attempt to prepare the cavity in the doubly excited state via two consecutive iSWAP gates. In this case, with only two mirror qubits and the rapid decay via the bright state of the two-excitation state $\Emirror$ of the mirror qubits (refer to Fig.~\ref{fig:Rabi}(e)), the resulting probe qubit population dynamics shown in Fig.~\ref{fig:Rabi}(f) have a strongly damped response ($\mathcal{C} <1$) with weak oscillations occurring at the vacuum Rabi oscillation frequency. This is in contrast to the behavior of a linear cavity (dashed green curve of Fig.~\ref{fig:Rabi}(f)), where driving the second photon transition leads to persistent Rabi oscillations with a frequency that is $\sqrt{2}$ larger than vacuum Rabi oscillations.  Further analysis of the nonlinear behavior of the atomic cavity is provided in App.~\ref{App:E}. 

The waveguide QED chip of Fig.~\ref{fig:schematic}(b) can also be used to investigate the spectrum of sub-radiant states that emerge when $N > 2$ and direct interaction between mirror qubits is manifest. This situation can be realized by taking advantage of the capacitive coupling between co-localized pairs of type-I qubits (Q$_2$ and Q$_3$ or Q$_5$ and Q$_6$). Although in an idealized 1D waveguide model there is no cooperative interaction term between qubits with zero separation along the waveguide, as shown in Fig.~\ref{fig:Compound}(a) we observe a strong coupling ($g/2\pi =  46$~MHz) between the co-localized pair of Q$_2$ and Q$_3$ mirror qubits. This coupling results from near-field components of the electromagnetic field that are excluded in the simple waveguide model. The non-degenerate hybridized eigenstates of the qubit pair effectively behave as a compound atomic mirror. The emission rate of each compound mirror to the waveguide can be adjusted by setting the detuning $\Delta$ between the pair. As illustrated in Fig.~\ref{fig:Compound}(b), resonantly aligning the compound atomic mirrors on both ends of the waveguide results in a hierarchy of bright and dark states involving both near-field and waveguide-mediated cooperative coupling. Probing the system with a weak tone via the waveguide, we identify the two super-radiant combinations of the compound atomic mirrors (Fig.~\ref{fig:Compound}(c)). Similar to the case of a two-qubit cavity, we can identify the collective dark states via the probe qubit.  As evidenced by the measured Rabi oscillations shown in Fig.~\ref{fig:Compound}(d), the combination of direct and waveguide-mediated interactions of mirror qubits in this geometry results in the emergence of a pair of collective entangled states of the four qubits acting as strongly-coupled atomic cavities with frequency separation of $\sqrt{4g^2+\Delta^2}$.


We envision several avenues of further exploration beyond the work presented here. We note that the measured frequency disorder of the qubits in our system ($\delta f \approx 60$ MHz) is smaller than the largest achieved qubit emission rate into the waveguide ($\Gamma_\text{1D}/2\pi \approx 100$ MHz).  A modest reduction in this $\delta f /\Gamma_\text{1D}$ ratio eliminates the need for frequency-tunable qubits, which can result in significant simplifications in circuit layout and superior qubit coherence~\cite{Hutchings:2017gc}.  Combining this with slow-light metamaterial waveguides~\cite{Mirhosseini:2018wg} would allow chip-scale waveguide QED experiments with a much larger number of qubits, in the range $N=10$--$100$, where the full extent of the rich physics of multi-excitation collective states can be studied. Examples include the formation of ``fermionic" spatial correlations in multi-qubit dark states~\cite{Albrecht:je}, deterministic generation of multi-photon entanglement~\cite{GonzalezTudela:2015ek,GonzalezTudela:2017kr}, and driven-dissipative generation of many-body quantum states of light under classical drive \cite{Ramos:2014,Ringel:2014,Mahmoodian:2018}. In addition, a crucial technical advancement in our current work is the achievement of single qubit Purcell factors in excess of $200$, a value more than an order of magnitude improved over previous work in planar superconducting quantum circuits and on par with the values achievable in less scalable 3D architectures~\cite{Hamann:2018tm}. Considering the state-of-the-art coherence times for planar superconducting qubits~\cite{Yan:2016df,Bronn:2018da}, and with improved thermalization of the waveguide~\cite{Yeh:2017bd,Wang:2018vz}, Purcell factors in excess of $10^4$ should be achievable.  In this regime, a universal set of quantum gates with fidelity above $0.99$ can theoretically be realized by encoding information in the decoherence-free subspaces (dark states) within the waveguide QED system~\cite{Paulisch:2016ib}. 


\begin{acknowledgments}
The authors thank Jen-Hao Yeh and Ben Palmer for the use of one of their cryogenic attenuators, critical to reducing thermal noise in the input waveguide line.  This work was supported by the AFOSR MURI Quantum Photonic Matter (grant FA9550-16-1-0323), the Institute for Quantum Information and Matter, an NSF Physics Frontiers Center (grant PHY-1125565) with support of the Gordon and Betty Moore Foundation, and the Kavli Nanoscience Institute at Caltech.  D.E.C. acknowledges support from the ERC Starting Grant FOQAL, MINECO Plan Nacional Grant CANS, MINECO Severo Ochoa Grant No. SEV-2015-0522, CERCA Programme/Generalitat de Catalunya, and Fundacio Privada.  M.M. is supported through a KNI Postdoctoral Fellowship, A.J.K. and A.S. are supported by IQIM Postdoctoral Scholarships, P.B.D. is supported by a Hertz Graduate Fellowship Award, and A.A.-G. is supported by the Global Marie Curie Fellowship under the LANTERN program.
\end{acknowledgments}


%

\appendix
\clearpage

\onecolumngrid

\section{Device Fundamentals}
\label{App:A}

\subsection{Fabrication}
The device used in this work is fabricated on a $1\:\textrm{cm}\times 1\:\textrm{cm}$ high resistivity 10\:k$\Omega$-cm silicon substrate. The ground plane, waveguides, resonator, and qubit capacitors are patterned by electron-beam lithography followed by electron beam evaporation of 120\:nm Al at a rate of 1 nm/s. A liftoff process is performed in N-methyl-2-pyrrolidone at 80\:$^{\circ}$C for 1.5 hours. The Josephson junctions are fabricated using double-angle electron-beam evaporation on suspended Dolan bridges, following similar techniques as in Ref.~\cite{Keller2017}. The airbridges are patterned using grayscale electron-beam lithography and developed in a mixture of isopropyl alcohol and deionized water \cite{Rooks2002}. After 2 hours of resist reflow at 105\:$^{\circ}$C, electron-beam evaporation of 140\:nm Al is performed at 1\:nm/s rate following 5 minutes of Ar ion mill. Liftoff is done in the same fashion as in the previous steps.

\begin{figure}[b]
\begin{center}
\includegraphics[width=\textwidth]{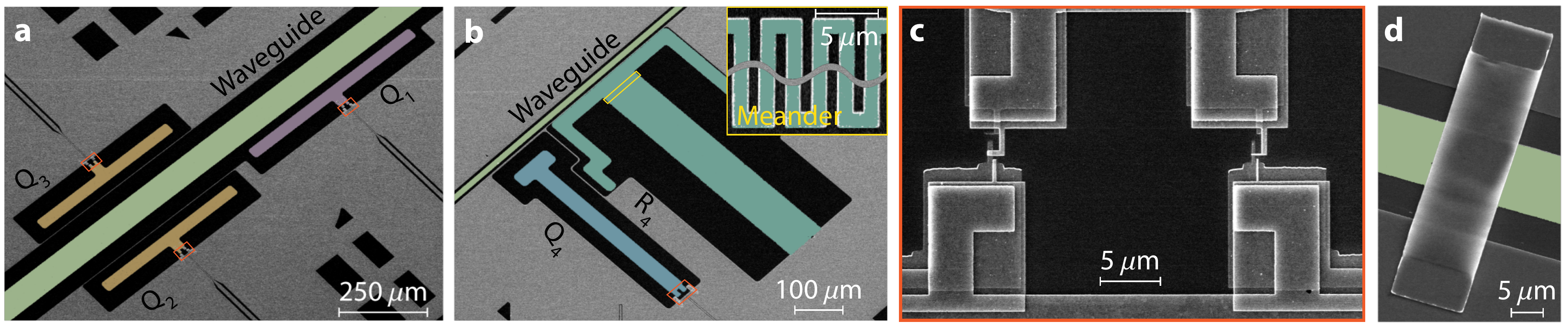}
\caption{\textbf{Scanning electron microscope of the fabricated device.} a. Type-I (Q$_2$, Q$_3$) and type-II (Q$_1$) mirror qubits coupled to the coplanar waveguide (CPW). b. The central probe qubit (Q$_4$) and lumped-element readout resonator (R$_4$) coupled to CPW. Inset: inductive meander of the lumped-element readout resonator. c. A superconducting quantum interference device (SQUID) loop with asymmetric Josephson junctions used for qubits. d. An airbridge placed across the waveguide to suppress slotline mode.} 
\label{fig:figureS1}
\end{center}
\end{figure}

\subsection{Qubit}
We have designed and fabricated transmon qubits in three different variants for the experiment: type-I mirror qubits (Q$_2$, Q$_3$, Q$_5$, Q$_6$), type-II mirror qubits (Q$_1$, Q$_7$), and the probe qubit (Q$_4$). The qubit frequency tuning range, waveguide coupling rate ($\Gamma_\text{1D}$), and parasitic decoherence rate ($\Gamma^\prime$) can be extracted from waveguide spectroscopy measurements of the individual qubits. The values for all the qubits inferred in this manner are listed in Table~\ref{tb:qubits}. Note that $\Gamma^\prime$ is defined as due to damping and dephasing from channels other than the waveguide at zero temperature.  The inferred value of $\Gamma^\prime$ from waveguide spectroscopy measurements is consistent with this definition in the zero temperature waveguide limit (effects of finite waveguide temperature are considered in Supplementary Information Note $2$).  The standard deviation in maximum frequencies of the four identically designed qubits (type-I) is found as $61\:\text{MHz}$, equivalent to $\sim1\%$ qubit frequency disorder in our fabrication process. Asymmetric Josephson junctions are used in all qubits' superconducting quantum interference device (SQUID) loops (Figure~\ref{fig:figureS1}c) to reduce dephasing from flux noise, which limits the tuning range of qubits to $\sim 1.3\:\text{GHz}$. For Q$_4$, the Josephson energy of the junctions are extracted to be $(E_{J1}, E_{J2})/h = (18.4,\: 3.5)\:\mathrm{GHz}$, giving junction asymmetry of $d\equiv\frac{E_{J1}-E_{J2}}{E_{J1}+E_{J2}}=0.68$. The anharmonicity was measured to be $\eta/2\pi = - 272\:\text{MHz}$ and $E_J/E_C= 81$ at maximum frequency for Q$_4$.
\begin{table}[h]
\begin{threeparttable}[t]
  \centering
  
\begin{tabular}{M{2.1cm}P{1.2cm}P{1.2cm}P{1.2cm}P{1.2cm}P{1.2cm}P{1.2cm}P{1.2cm}} 
               & Q$_1$    & Q$_2$    & Q$_3$    & Q$_4$    & Q$_5$    & Q$_6$    & Q$_7$    \\ \hline \hline
$f_\text{max}$ (GHz) & 6.052 & 6.678 & 6.750 & 6.638 & 6.702 & 6.817  & 6.175 \\
$f_\text{min}$ (GHz)  & 4.861 & 5.373 & 5.389 & 5.431 & 5.157 & 5.510 & 4.972 \\  \hline
$\Gamma_\text{1D}/2\pi$  (MHz) &  94.1     &   16.5    &    13.9\tnote{a,b}   &    0.91   &    18.4\tnote{b}   &    18.1   &   99.5   \\
$\Gamma^\prime/2\pi$  (kHz)  &    430   &    $<341$   &  $<760$\tnote{a,b}   &    81   &     375\tnote{b}  &    185   &  998
 \end{tabular}
     \begin{tablenotes}
     \item[a] Measured at 6.6\:GHz
     \item[b] Measured without the cold attenuator
   \end{tablenotes}
   \end{threeparttable}
 \caption{\textbf{Qubit characteristics.} $f_\text{max}$ ($f_\text{min}$) is the maximum (minimum) frequency of the qubit, corresponding to ``sweet spots" with zero first-order flux sensitivity. $\Gamma_\text{1D}$ is the qubit's rate of decay into the waveguide channel and $\Gamma^\prime$ is its parasitic decoherence rate due to damping and dephasing from channels other than the waveguide at 0 temperature.  All reported values are measured at the maximum frequency of each qubit, save for Q$_3$ in which case the values were measured at $6.6$~GHz (marked with superscript $^\mathrm{a}$). With the exception of Q$_3$ and Q$_5$ (marked with superscript $^\mathrm{b}$), all the values are measured with the cold attenuator placed in the input line of the waveguide (see App.~\ref{App:B}).}
   \label{tb:qubits}
\end{table}

\subsection{Readout}
We have fabricated a lumped-element resonator (shown in Fig.~\ref{fig:figureS1}b) to perform dispersive readout of the state of central probe qubit (Q$_4$). The lumped-element resonator consists of a capacitive claw and an inductive meander of $\sim 1\:\mu\text{m}$ pitch, effectively acting as a quarter-wave resonator. The bare frequency of resonator and coupling to probe qubit are extracted to be $f_\text{r}=5.156\:\textrm{GHz}$ and $g/2\pi = 116\:\textrm{MHz}$, respectively, giving dispersive frequency shift of $\chi/2\pi = -2.05\:\textrm{MHz}$ for Q$_4$ at maximum frequency. The resonator is loaded to the common waveguide in the experiment, and its internal and external quality factors are measured to be $Q_i = 1.3\times 10^{5}$ and $Q_e= 980$ below single-photon level. It should be noted that the resonator-induced Purcell decay rate of Q$_4$ is $\Gamma_{1}^\text{Purcell}/ 2\pi \sim 70\:\textrm{kHz}$, small compared to the decay rate into the waveguide $\Gamma_\mathrm{1D,p}/2\pi\sim 1\:\text{MHz}$. The compact footprint of the lumped-element resonator is critical for minimizing the distributed coupling effects that may arise from interference between direct qubit decay to the waveguide and the the Purcell decay of the qubit via the resonator path.
\subsection{Airbridge}
In our experiment we use a coplanar transmission line for realizing a microwave waveguide. In addition to the fundamental propagating mode of the waveguide, which has even symmetry with respect to the waveguide axis, these structures also support a set of modes with the odd symmetry, known as the slotline modes. The propagation of the slotline mode can be completely suppressed in a waveguide with perfectly symmetric boundary conditions. However, in practice perfect symmetry cannot be maintained over the full waveguide length, which unavoidably leads to presence of the slotline mode as a spurious loss channel for the qubits. Crossovers connecting ground planes across the waveguide are known to suppress propagation of slotline mode, and to this effect, aluminum airbridges have been used in superconducting circuits with negligible impedance mismatch for the desired CPW mode ~\cite{Chen2013}.

In this experiment, we place airbridges (Fig.~\ref{fig:figureS1}d) along the waveguide and control lines with the following considerations. Airbridges create reflecting boundary for slotline mode, and if placed by a distance $d$ a discrete resonance corresponding to wavelength of $2d$ is formed.  By placing airbridges over distances smaller than $\lambda/4$ apart from each other ($\lambda$ is the wavelength of the mode resonant with the qubits), we push the slotline resonances of the waveguide sections between the airbridges to substantially higher frequencies. In this situation, the dissipation rate of qubits via the spurious channel is significantly suppressed by the off-resonance Purcell factor $\Gamma_1^\text{Purcell}\sim (g/\Delta)^2\kappa$, where $\Delta$ denotes detuning between the qubit transition frequency and the frequency of the odd mode in the waveguide section between the two airbridges. The parameters $g$ and $\kappa$ are the interaction rate of the qubit and the decay rate of the slot-line cavity modes. In addition, we place the airbridges before and after bends in waveguide, to ensure the fundamental waveguide mode is not converted to the slot-line mode upon propagation~\cite{Ming-DongWu1995}.

\subsection{Flux Crosstalk}
We tune the frequency of each qubit by supplying a bias current to individual Z control lines, which controls the magnetic flux in the qubit's SQUID loop. The bias currents are generated via independent bias voltages generated by seven arbitrary waveform generator (AWG) channels, allowing for simultaneous tuning of all qubits. In practice, independent frequency tuning of each qubit needs to be accompanied by small changes in the flux bias of the qubits in the near physical vicinity of the qubit of interest, due to cross-talk between adjacent Z control lines.

In this experiment, we have characterized the crosstalk between bias voltage channels of the qubits in the following way. First, we tune the qubits not in use to frequencies more than $800\:\text{MHz}$ away from the working frequency (which is set as either 5.83\:GHz or 6.6\:GHz). These qubits are controlled by fixed biases such that their frequencies, even in the presence of crosstalk from other qubits, remain far enough from the working frequency and hence are not considered for the rest of the analysis. Second, we tune the remaining qubits in use to relevant frequencies within $100\:\mathrm{MHz}$ of the working frequency and record the biases $\mathbf{v}_0$ and frequencies $\mathbf{f}_0$ of these qubits. Third, we vary the bias on only a single ($j$-th) qubit and linearly interpolate the change in frequency ($f_i$) of the other ($i$-th) qubits with respect to bias voltage $v_j$ on $j$-th qubit, finding the cross talk matrix component $M_{ij} = (\partial f_{i}/\partial v_{j})_{\mathbf{v}=\mathbf{v}_0}$. Repeating this step, we get the following (approximately linearized) relation between frequencies $\mathbf{f}$ and bias voltages $\mathbf{v}$ of qubits:
\begin{equation}
    \mathbf{f} \approx \mathbf{f}_0 + M (\mathbf{v} - \mathbf{v}_0).\label{eq:crosstalk-relation}
\end{equation}
Finally, we take the inverse of relation \eqref{eq:crosstalk-relation} to find bias voltages $\mathbf{v}$ that is required for tuning qubits to frequencies $\mathbf{f}$:
$$\mathbf{v} \approx \mathbf{v}_0 + M^{-1} (\mathbf{f} - \mathbf{f}_0). $$
An example of such crosstalk matrix between Q$_2$, Q$_4$, and Q$_6$ near $\mathbf{f}_0 = (6.6, 6.6, 6.6)\:\text{GHz}$ used in the experiment is given by
$$M =
\begin{pmatrix} 
    0.2683   &  -0.0245 &  -0.0033\\
   -0.0141   & -0.5310  &  0.0170\\
    0.0016   &  0.0245  &  0.4933\\
\end{pmatrix} \mathrm{{GHz}/V}$$
This indicates that the crosstalk level between Q$_4$ and either Q$_2$ or Q$_6$ is about 5\%, while that between Q$_2$ and Q$_6$ is less than 1\%. We have repeated similar steps for other configurations in the experiment.

\subsection{Experimental Setup}
\begin{figure}[b]
\begin{center}
\includegraphics[width=\textwidth]{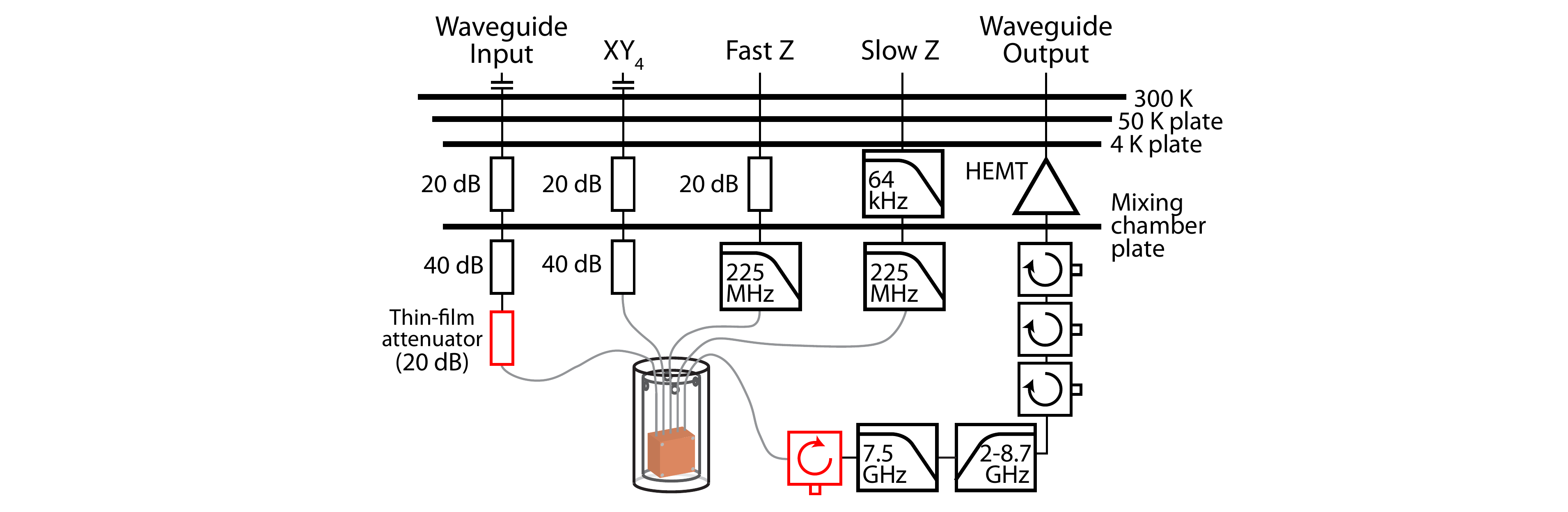}
\caption{\textbf{Schematic of the measurement chain inside the dilution refrigerator.} The four types of input lines, the output line, and their connection to the device inside a magnetic shield are illustrated. Attenuators are expressed as rectangles with labeled power attenuation and capacitor symbols correspond to DC blocks. The thin-film attenuator and a circulator (colored red) are added to the waveguide input line and output line, respectively, in a second version of the setup and a second round of measurements to further protect the sample from thermal noise in the waveguide line.} 
\label{fig:figureS2}
\end{center}
\end{figure}

Figure~\ref{fig:figureS2} illustrates the outline of the measurement chain in our dilution refrigerator. The sample is enclosed in a magnetic shield which is mounted at the mixing chamber. We have outlined four different types of input lines used in our experiment. Input lines to the waveguide and XY$_4$ go through a DC block at room temperature and are attenuated by 20\:dB at the 4\:K stage, followed by additional 40\:dB of attenuation at the mixing chamber. The fast flux tuning lines (Z$_3$, Z$_4$) are attenuated by 20 dB and are filtered with a low-pass filter with corner frequency at $225$~MHz to minimize thermal noise photons while maintaining short rise and fall time of pulses for fast flux control. The slow flux tuning lines (Z$_1$, Z$_2$, Z$_5$, Z$_6$, Z$_7$) are filtered by an additional low-pass filter with $64$~kHz corner frequency at the 4K stage to further suppress noise photons. In addition, the waveguide signal output path contains a high electron mobility transistor (HEMT) amplifier at the $4$K plate. Three circulators are placed in between the HEMT and the sample to ensure ($>70$\:dB) isolation of the sample from the amplifier noise. In addition, we have a series of low-pass and band-pass filters on the output line to suppress noise sources outside the measurement spectrum.

A thin-film ``cold attenuator", developed in Ben Palmer's group at the University of Maryland~\cite{Yeh:2017bd} to better thermalize a microwave coaxial line to its environment, as well as an additional circulator (both attenuator and circulator are highlighted in red in the schematic), are added to the waveguide measurement chain in later setups to further protect the device against thermal photons. The effect of this change is discussed in App.~\ref{App:B}.

\section{Spectroscopic measurement of individual qubits}
\label{App:B}
The master equation of a qubit in a thermal bath at temperature $T$, driven by a classical field is given by $\dot{\hat{\rho}} = -i[\hat{H}/\hbar, \hat{\rho}] + \mathcal{L}[\hat{\rho}]$,
where the Hamiltonian $\hat{H}$ and the Liouvillian $\mathcal{L}$ is written as \cite{Gardiner2004}
\begin{align}
    \hat{H}/\hbar &= -\frac{\omegap - \omegaq}{2}\: \hat{\sigma}_z + \frac{\Omegap}{2}\:\hat{\sigma}_x ,\\
    \mathcal{L}[\hat{\rho}] &= (\nth + 1)\Gamma_{1} \mathcal{D}[\hat{\sigma}_-]\hat{\rho} +  \nth \Gamma_{1} \mathcal{D}[\hat{\sigma}_+]\hat{\rho} + \frac{\Gammaphi}{2}\mathcal{D}[\hat{\sigma}_z]\hat{\rho}.
\end{align}
Here, $\omegap$ ($\omegaq$) is the frequency of the drive (qubit), $\Omegap$ is the Rabi frequency of the drive, $\nth = 1/(e^{\hbar\omegaq/\kB T} - 1)$ is the thermal occupation of photons in the bath, $\Gamma_1$ and $\Gammaphi$ are relaxation rate and pure dephasing rates of the qubit, respectively. The superoperator
\begin{equation}
\mathcal{D}[\hat{A}]\hat{\rho} = \hat{A}\hat{\rho}\hat{A}^\dagger - \frac{1}{2}\{\hat{A}^\dagger\hat{A}, \hat{\rho}\} \label{eq:lindblad-dissipator}
\end{equation}
denotes the Lindblad dissipator. The master equation can be rewritten in terms of density matrix elements $\rho_{a,b}\equiv\langle a|\hat{\rho} | b\rangle$ as
\begin{align}
    \drhoee &= \frac{i\Omegap}{2}(\rhoeg - \rhoge) - (\nth + 1)\Gamma_1 \rhoee + \nth \Gamma_1 \rhogg \\
    \drhoeg &= \left[i(\omegap - \omegaq) - \frac{(2\nth + 1)\Gamma_1 + 2\Gammaphi}{2}\right]\rhoeg + \frac{i\Omegap}{2}(\rhoee - \rhogg) \\
    \drhoge &= \drhoeg^* ;\quad \drhogg= -\drhoee
\end{align}
With $\rhoee + \rhogg = 1$, the steady-state solution ($\dot{\hat{\rho}}=0$) to the master equation can be expressed as
\begin{align}
    \rhoeess &= \frac{\nth}{2\nth + 1} \frac{1 + (\domega/\Gammatwoth)^2 }{1 + (\domega/\Gammatwoth)^2 + \Omegap^2/(\Gammaoneth \Gammatwoth)} + \frac{1}{2} \frac{\Omegap^2 / (\Gammaoneth \Gammatwoth)}{1 + (\domega/\Gammatwoth)^2 + \Omegap^2/(\Gammaoneth \Gammatwoth)},\\
    \rhoegss &= -i \frac{\Omegap}{2\Gammatwoth(2\nth + 1)}\frac{1 + i\:\domega/\Gammatwoth}{1 + (\domega/\Gammatwoth)^2 + \Omegap^2/(\Gammaoneth \Gammatwoth)},\label{eq:rhoegss}
\end{align}
where $\delta\omega = \omegap - \omegaq$ is the detuning of the drive from qubit frequency, $\Gammaoneth=(2\nth + 1)\Gamma_1$ and $\Gammatwoth = \Gammaoneth/2 + \Gammaphi$ are the thermally enhanced decay rate and dephasing rate of the qubit.

Now, let us consider the case where a qubit is coupled to the waveguide with decay rate of $\Gamma_\mathrm{1D}$. If we send in a probe field $\ain$ from left to right along the waveguide, the right-propagating output field $\aout$ after interaction with the qubit is written as \cite{Lalumiere:2013io}
$$\aout = \ain + \sqrt{\frac{\GammaoneD}{2}}\sigmam.$$
The probe field creates a classical drive on the qubit with the rate of $\Omegap/2=-i\langle\ain\rangle\sqrt{\Gamma_\text{1D}/2}$. 
With the steady-state solution of master equation \eqref{eq:rhoegss} the transmission amplitude $t=\langle\aout\rangle/\langle\ain\rangle$ can be written as 
\begin{equation}
t(\domega) = 1 - \frac{\GammaoneD}{2\Gammatwoth(2\nth + 1)}\frac{1 + i\:\domega/\Gammatwoth}{1 + (\domega/\Gammatwoth)^2 + \Omegap^2/(\Gammaoneth \Gammatwoth)}.\label{eq:transmission-thermal}
\end{equation}
At zero temperature ($\nth=0$) Eq.~\eqref{eq:transmission-thermal} reduces to~\cite{Astafiev:2010cm, Peropadre2013},
\begin{equation}
t(\domega) = 1 - \frac{\GammaoneD}{2\Gammatwo}\frac{1 + i\:\domega/\Gammatwo}{1 + (\domega/\Gammatwo)^2 + \Omegap^2/(\Gammaone \Gammatwo)}.\label{eq:transmission-vacuum}
\end{equation}
Here, $\Gammatwo=\Gammaphi + \Gammaone/  2$ is the dephasing rate of the qubit in the absence of thermal occupancy. In the following, we define the parasitic decoherence rate of the qubit as $\Gamma^\prime = 2\Gamma_2 - \GammaoneD = \Gammaloss + 2\Gammaphi$, where $\Gammaloss$ denotes the decay rate of qubit induced by channels other than the waveguide. Examples of $\Gammaloss$ in superconducting qubits include dielectric loss, decay into slotline mode, and loss from coupling to two-level system (TLS) defects.
\subsection{Effect of saturation}
To discuss the effect of saturation on the extinction in transmission, we start with the zero temperature case of Eq.~\eqref{eq:transmission-vacuum}. We introduce the saturation parameter $s\equiv \Omegap^2/ \Gammaone\Gammatwo$ to rewrite the on-resonance transmittivity as
\begin{equation}
t(0) = 1 - \frac{\Gamma_\mathrm{1D}}{2\Gamma_2} \frac{1}{1+s} \approx 1 - \frac{\Gamma_\mathrm{1D}}{2\Gamma_2}(1-s) = \left(1 + s\frac{\Gamma_\mathrm{1D}}{\Gamma^\prime} \right) \left( \frac{\Gamma^\prime}{\Gamma^\prime + \Gamma_\mathrm{1D}}\right), \label{eq:saturation}
\end{equation}
where the low-power assumption $s \ll 1$ has been made in the last step. For the extinction to get negligible effect from saturation, the power-dependent part in Eq.~\eqref{eq:saturation} should be small compared to the power-independent part. This is equivalent to $s< \Gamma^\prime / \Gamma_\mathrm{1D}$. Using the relation $$\Omegap = \sqrt{\frac{2\Gamma_\mathrm{1D} \Pp}{\hbar\omega_\mathrm{q}}}$$ between the driven Rabi frequency and the power $\Pp$ of the probe and assuming $\Gammaprime \ll \GammaoneD$, this reduces to 
\begin{equation}
    \Pp \lesssim \frac{\hbar\omega_\mathrm{q}\Gamma^\prime}{4}.
\end{equation}
In the experiment, the probe power used to resolve the extinction was -150~dBm ($10^{-18}\:\text{W}$), which gives a limit to the observable $\Gammaprime$ due to our coherent drive of $\Gammaprime/2\pi \approx 150\:\text{kHz}$.
\subsection{Effect of thermal occupation}
\begin{figure}[b]
\begin{center}
\includegraphics[width=0.5\textwidth]{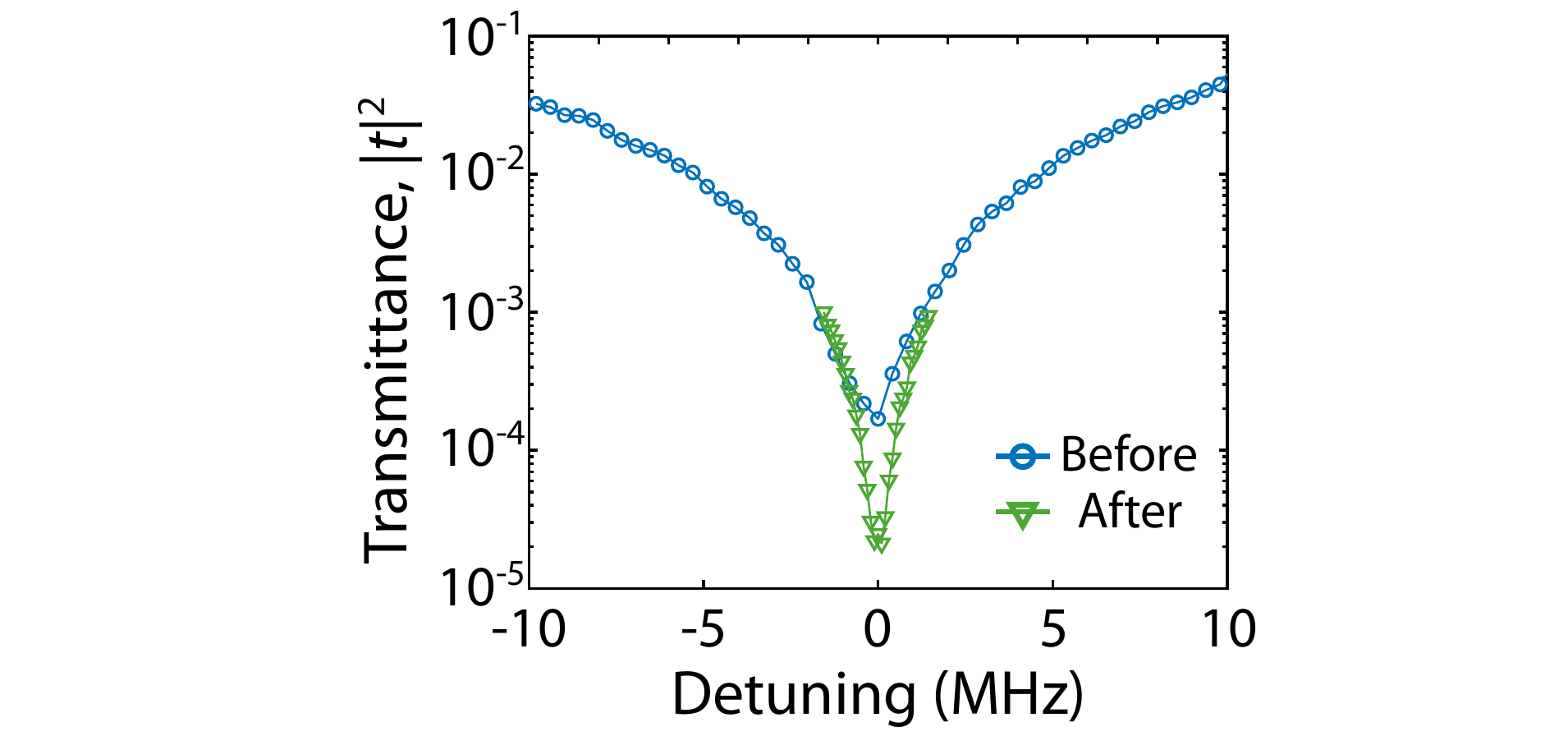}
\caption{\textbf{Effect of thermal occupancy on extinction.} The transmittance of Q$_1$ is measured at the flux-insensitive point before and after installation of customized microwave attenuator. We observe an order-of-magnitude enhancement in extinction after the installation, indicating a better thermalization of input signals to the chip.} 
\label{fig:figureS3}
\end{center}
\end{figure}
To take into account the effect of thermal occupancy, we take the limit where the saturation is very small $(\Omegap\approx 0)$. On resonance, the transmission amplitude is expressed as 
\begin{equation}
t(0) = 1 - \frac{\GammaoneD}{[(2\nth + 1)\Gammaone + 2\Gammaphi](2\nth + 1))} \approx 1- \frac{\GammaoneD}{2\Gammatwo} + \frac{(\Gammaone + \Gammaphi)\GammaoneD}{\Gammatwo^2 }\nth,
\end{equation}
where we have assumed the thermal occupation is very small, $\nth\ll 1$. In the limit where $\GammaoneD$ is dominating spurious loss and pure dephasing rates ($\Gammatwo\approx \GammaoneD/2$), this reduces to
\begin{equation}
    t(0) \approx t(0)|_{T=0} + 4\nth
\end{equation}
and hence the thermal contribution dominates the transmission amplitude unless $\nth < \Gammaprime / 4\GammaoneD$.

Using this relation, we can estimate the upper bound on the temperature of the environment based on our measurement of extinction. We have measured the transmittance of Q$_1$ at its maximum frequency (Fig.~\ref{fig:figureS3}) before and after installing a thin-film microwave attenuator, which is customized for proper thermalization of the input signals sent into the waveguide with the mixing chamber plate of the dilution refrigerator~\cite{Yeh:2017bd}. The minimum transmittance was measured to be $|t|^2\approx 1.7\times 10^{-4}\ (2.1\times 10^{-5})$ before (after) installation of the attenuator, corresponding to the upper bound on thermal photon number of $\nth \lesssim 3.3\times 10^{-3}\ (1.1\times 10^{-3}$). With the attenuator, this corresponds to temperature of 43\:mK, close to the temperature values reported in Ref.~\cite{Yeh:2017bd}.

\section{Detailed modeling of the atomic cavity}
\label{App:C}

\begin{figure}[b]
\begin{center}
\includegraphics[width=0.5\textwidth]{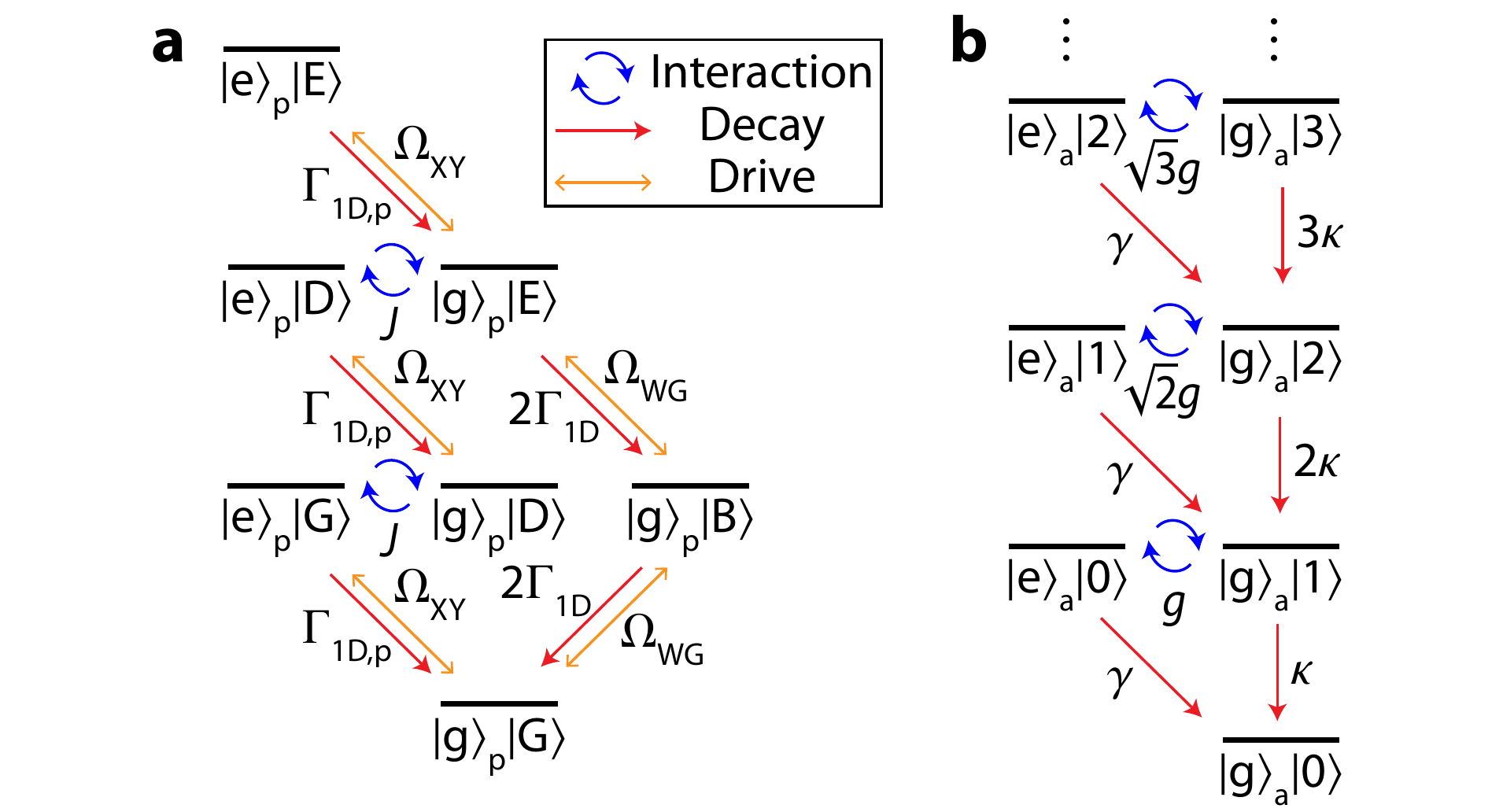}
\caption{\textbf{Level structure of the atomic cavity and linear cavity.} a. Level structure of the three-qubit system of probe qubit and atomic cavity. $\GammaoneDp$ and $2\GammaoneD$ denotes the decay rates into the waveguide channel, $\Omega_\text{XY}$ is the local drive on the probe qubit, and $\Omega_\text{WG}$ is the drive from the waveguide. The coupling strength $J$ is the same for the first excitation and second excitation levels, b. Level structure of an atom coupled to a linear cavity. $|e\rangle_\text{a}$ ($|g\rangle_\text{a}$) denotes the excited state (ground state) of the atom, while $|n\rangle$ is the $n$-photon Fock state of the cavity field. $g$ is the coupling, $\gamma$ is the decay rate of the atom, and $\kappa$ is the photon loss rate of the cavity.} 
\label{fig:figureS4}
\end{center}
\end{figure}

In this section, we analyze the atomic cavity discussed in the main text in more detail, taking into account its higher excitation levels. The atomic cavity is formed by two identical \emph{mirror} qubits [frequency $\omegaq$, decay rate $\GammaoneD$ ($\Gammaprime$) to waveguide (spurious loss) channel placed at $\lambda/2$ distance along the waveguide (Figure~1a). From the $\lambda/2$ spacing, the correlated decay of the two qubits is maximized to $-\GammaoneD$, while the exchange interaction is zero. This results in formation of dark state $|\text{D}\rangle$ and bright state $|\text{B}\rangle$
\begin{equation}
    |\text{D}\rangle = \frac{|\text{eg}\rangle + |\text{ge}\rangle}{\sqrt{2}},\quad |\text{B}\rangle = \frac{|\text{eg}\rangle - |\text{ge}\rangle}{\sqrt{2}}, \label{eq:DarkBrightStates}
\end{equation}
which are single-excitation states of two qubits with suppressed and enhanced waveguide decay rates $\Gamma_\text{1D,D}=0$, $\Gamma_\text{1D,B}=2\GammaoneD$ to the waveguide. Here, g (e) denotes the ground (excited) state of each qubit. Other than the ground state $|\text{G}\rangle\equiv |\text{gg}\rangle$, there also exists a second excited state $|\text{E}\rangle\equiv |\text{ee}\rangle$ of two qubits, completing $2^2=4$ eigenstates in the Hilbert space of two qubits.
We can alternatively define $|\text{D}\rangle$ and $|\text{B}\rangle$ in terms of collective annihilation operators
\begin{equation}
    \hat{S}_\text{D} = \frac{1}{\sqrt{2}}\left(\sigmam^{(1)} + \sigmam^{(2)}\right),\quad
 \hat{S}_\text{B} = \frac{1}{\sqrt{2}}\left(\sigmam^{(1)} - \sigmam^{(2)}\right) \label{eq:BrightDarkSigmaSymm}
\end{equation}
as $|\text{D}\rangle = \hat{S}_\text{D}^\dagger |\text{G}\rangle$ and  $|\text{B}\rangle = \hat{S}_\text{B}^\dagger |\text{G}\rangle$. Here, $\sigmam^{(i)}$ de-excites the state of $i$-th mirror qubit. Note that the doubly-excited state $|\text{E}\rangle$ can be obtained by successive application of either $\hat{S}_\text{D}^\dagger$ or $\hat{S}_\text{B}^\dagger$ twice on the ground state $|\text{G}\rangle$.

The interaction of qubits with the field in the waveguide is written in the form of $\hat{H}_\text{WG}\propto (\hat{S}_\text{B} + \hat{S}_\text{B}^\dagger),$ and hence the state transfer via classical drive on the waveguide can be achieved only between states of non-vanishing transition dipole $\langle f|\hat{S}_\text{B}|i\rangle $. In the present case, only $|\text{G}\rangle\leftrightarrow|\text{B}\rangle$ and $|\text{B}\rangle\leftrightarrow|\text{E}\rangle$ transitions are available via the waveguide with the same transition dipole. This implies that the waveguide decay rate of $|\text{E}\rangle$ is equal to that of $|\text{B}\rangle$, $\Gamma_\text{1D,E}=2\GammaoneD$. 

To investigate the level structure of the dark state, which is not accessible via the waveguide channel, we introduce an ancilla \emph{probe} qubit [frequency $\omegaq$, decay rate $\GammaoneDp$ ($\Gammaprime_\text{p}$) to waveguide (loss) channel] at the center of mirror qubits. The probe qubit is separated by $\lambda/4$ from mirror qubits, maximizing the exchange interaction to $\sqrt{\GammaoneDp\GammaoneD}/2$ with zero correlated decay. This creates an interaction of excited state of probe qubit to the dark state of mirror qubits $|\text{e}\rangle_\text{p}|\text{G}\rangle \leftrightarrow |\text{g}\rangle_\text{p}|\text{D}\rangle$, while the bright state remains decoupled from this dynamics.

The master equation of the three-qubit system reads  $\dot{\hat{\rho}} = -\frac{i}{\hbar}[\hat{H},\hat{\rho}] + \mathcal{L}[\hat{\rho}]$, where the Hamiltonian $\hat{H}$ and the Liouvillian $\mathcal{L}$ are given by
\begin{align}
    \hat{H} &= \hbar J\left[\sigmam^{\text{(p)}}\hat{S}_\text{D}^\dagger + \sigmap^{\text{(p)}}\hat{S}_\text{D} \right]\\
    \mathcal{L}[\hat{\rho}] &= (\GammaoneDp + \Gammaprime_\text{p}) \:\mathcal{D}\left[\sigmam^\text{(p)}\right]\hat{\rho} + (2\GammaoneD + \Gammaprime)\: \mathcal{D}\left[\hat{S}_\text{B}\right]\hat{\rho} + \Gammaprime\: \mathcal{D}\left[\hat{S}_\text{D}\right]\hat{\rho} \label{eq:Liouvillian-incoherent}
\end{align}
Here, $\hat{\sigma}_\pm^{\text{(p)}}$ are the Pauli operators for the probe qubit, $2J=\sqrt{2\GammaoneDp\GammaoneD}$ is the interaction between probe qubit and dark state, and $\mathcal{D}[\cdot]$ is the Lindblad dissipator defined in Eq.~\eqref{eq:lindblad-dissipator}. The full level structure of the $2^3=8$ states of three qubits and the rates in the system are summarized in Fig.~\ref{fig:figureS4}a. Note that the effective (non-Hermitian) Hamiltonian $\hat{H}_\text{eff}$ in the main text can be obtained from absorbing part of the Liouvillian in Eq.~\eqref{eq:Liouvillian-incoherent} excluding terms associated with quantum jumps. 

To reach the dark state of the atomic cavity, we first apply a local gate $|\text{g}\rangle_\text{p}|\text{G}\rangle\rightarrow |\text{e}\rangle_\text{p}|\text{G}\rangle$ on the probe qubit ($\Omega_\text{XY}$ in Fig.~\ref{fig:figureS4}a) to prepare the state in the first-excitation manifold. Then, the Rabi oscillation $|\text{e}\rangle_\text{p}|\text{G}\rangle\leftrightarrow |\text{g}\rangle_\text{p}|\text{D}\rangle$ takes place with the rate of $J$. We can identify $g=J$, $\gamma=\GammaoneDp + \Gammaprime_\text{p}$, $\kappa=\Gammaprime$ in analogy to cavity QED (Fig.~\ref{fig:schematic}a and Fig.~\ref{fig:figureS4}b) and calculate cooperativity as
$$
\mathcal{C} = \frac{(2J)^2}{\Gamma_{1,\text{p}}\Gamma_{1,\text{D}}}=\frac{2\GammaoneDp\GammaoneD}{(\GammaoneDp + \Gammaprime_\text{p})\Gammaprime} \approx \frac{2\GammaoneD}{\Gammaprime},
$$
when the spurious loss rate $\Gammaprime$ is small. A high cooperativity can be achieved in this case due to collective suppression of radiation in atomic cavity and cooperative enhancement in the interaction, scaling linearly with the Purcell factor $P_\text{1D}=\GammaoneD/\Gammaprime$. Thus, we can successfully map the population from the excited state of probe qubit to dark state of mirror qubits with the interaction time of $(2J/\pi)^{-1}$.

Going further, we attempt to reach the second-excited state $|\text{E}\rangle=(\hat{S}_\text{D}^\dagger)^2|\text{G}\rangle$ of atomic cavity. After the state preparation of $|\text{g}\rangle_\text{p}|\text{D}\rangle$ mentioned above, we apply another local gate $|\text{g}\rangle_\text{p}|\text{D}\rangle\rightarrow |\text{e}\rangle_\text{p}|\text{D}\rangle$ on the probe qubit and prepare the state in the second-excitation manifold.
In this case, the second excited states $|\text{e}\rangle_\text{p}|\text{D}\rangle\leftrightarrow |\text{g}\rangle_\text{p}|\text{E}\rangle$ have interaction strength $J$, same as the first excitation, while the $|\text{E}\rangle$ state becomes highly radiative to waveguide channel. The cooperativity $\mathcal{C}$ is calculated as
$$
\mathcal{C} = \frac{(2J)^2}{\Gamma_{1,\text{p}}\Gamma_{1,\text{E}}} = \frac{2\GammaoneDp\GammaoneD}{(\GammaoneDp + \Gammaprime_\text{p})(2\GammaoneD + \Gammaprime)} < 1,
$$
which is always smaller than unity. Therefore, the state $|\text{g}\rangle_\text{p}|\text{E}\rangle$
 is only virtually populated and the interaction maps the population in $|\text{e}\rangle_\text{p}|\text{D}\rangle$ to $|\text{g}\rangle_\text{p}|\text{B}\rangle$ with the rate of $(2J)^2/(2\GammaoneD)= \GammaoneDp$. This process competes with radiative decay (at a rate of $\GammaoneDp$) of probe qubit $|\text{e}\rangle_\text{p}|\text{D}\rangle \rightarrow |\text{g}\rangle_\text{p}|\text{D}\rangle$ followed by the Rabi oscillation in the first-excitation manifold, giving rise to damped Rabi oscillation in Fig.~3f. 

\subsection{Effect of phase length mismatch}
Deviation of phase length between mirror qubits from $\lambda/2$ along the waveguide can act as a non-ideal contribution in the dynamics of atomic cavity. The waveguide decay rate of dark state can be written as $\Gamma_\text{1D,D} = \GammaoneD ( 1 - |\cos{\phi}|)$, where $\phi=k_\text{1D} d$ is the phase separation between mirror qubits \cite{Lalumiere:2013io}. Here, $k_\text{1D}$ is the wavenumber and $d$ is the distance between mirror qubits.

We consider the case where the phase mismatch $\Delta\phi = \phi - \pi$ of mirror qubits is small. The decay rate of the dark state scales as $\Gamma_\text{1D,D}\approx\GammaoneD(\Delta\phi)^2/2$ only adding a small contribution to the decay rate of dark state. Based on the decay rate of dark states from time-domain measurement in Table~\ref{tb:darkDynamics}, we estimate the upper bound on the phase mismatch $\Delta\phi/\pi$ to be 5\% for type-I and 3.5\% for type-II.


\subsection{Effect of asymmetry in $\GammaoneD$}
So far we have assumed that the waveguide decay rate $\GammaoneD$ of mirror qubits are identical and neglected the asymmetry. If the waveguide decay rates of mirror qubits are given by $\Gamma_\text{1D,1} \neq \Gamma_\text{1D,2}$, the dark state and bright state are redefined as
\begin{equation}
    |\text{D}\rangle = \frac{\sqrt{\Gamma_\text{1D,2}}|\text{eg}\rangle + \sqrt{\Gamma_\text{1D,1}}|\text{ge}\rangle}{\sqrt{\Gamma_\text{1D,1}+\Gamma_\text{1D,2}}},\quad |\text{B}\rangle = \frac{\sqrt{\Gamma_\text{1D,1}}|\text{eg}\rangle - \sqrt{\Gamma_\text{1D,2}}|\text{ge}\rangle}{\sqrt{\Gamma_\text{1D,1}+\Gamma_\text{1D,2}}}, \label{eq:DarkBrightStatesAsymmetric}
\end{equation}
with collectively suppressed and enhanced waveguide decay rates of $\Gamma_\text{1D,D} =0$, $\Gamma_\text{1D,D} =\Gamma_\text{1D,1}+\Gamma_\text{1D,2}$, remaining fully dark and fully bright even in the presence of asymmetry. We also generalize Eq.~\eqref{eq:BrightDarkSigmaSymm} as
\begin{equation}
    \hat{S}_\text{D} = \frac{\sqrt{\Gamma_\text{1D,2}}\sigmam^{(1)} +\sqrt{\Gamma_\text{1D,1}} \sigmam^{(2)}}{\sqrt{\Gamma_\text{1D,1}+\Gamma_\text{1D,2}}},\quad
 \hat{S}_\text{B} = \frac{\sqrt{\Gamma_\text{1D,1}}\sigmam^{(1)} -\sqrt{\Gamma_\text{1D,2}} \sigmam^{(2)}}{\sqrt{\Gamma_\text{1D,1}+\Gamma_\text{1D,2}}}.
\end{equation}
With this basis, the Hamiltonian can be written as
\begin{equation}
    \hat{H} = \hbar J_\text{D}\left(\sigmam^\text{(p)} \hat{S}_\text{D}^\dagger + \sigmap^\text{(p)} \hat{S}_\text{D} \right)+ \hbar J_\text{B}\left(\sigmam^\text{(p)} \hat{S}_\text{B}^\dagger+\sigmap^\text{(p)} \hat{S}_\text{B} \right),
\end{equation}
where
$$
J_\text{D} = \frac{\sqrt{\GammaoneDp \Gamma_\text{1D,1} \Gamma_\text{1D,2}}}{\sqrt{\Gamma_\text{1D,1} + \Gamma_\text{1D,2}}}, \quad J_\text{B} = \frac{\sqrt{\GammaoneDp}( \Gamma_\text{1D,1} -\Gamma_\text{1D,2})}{2\sqrt{\Gamma_\text{1D,1} + \Gamma_\text{1D,2}}}.
$$
Thus, the probe qubit interacts with both the dark state and bright state with the ratio of $J_\text{D}:J_\text{B}=2\sqrt{\Gamma_\text{1D,1} \Gamma_\text{1D,2}}:(\Gamma_\text{1D,1} -\Gamma_\text{1D,2})$, and thus for a small asymmetry in the waveguide decay rate, the coupling to the dark state dominates the dynamics. In addition, we note that the bright state superradiantly decays to the waveguide, and it follows that coupling of probe qubit to the bright state manifest only as contribution of $$\frac{(2J_\text{B})^2}{\Gamma_\text{1D,1} +\Gamma_\text{1D,2}} = \GammaoneDp \left(\frac{\Gamma_\text{1D,1} -\Gamma_\text{1D,2}}{\Gamma_\text{1D,1} +\Gamma_\text{1D,2}}\right)^2$$
 to the probe qubit decay rate into spurious loss channel. In our experiment, the maximum asymmetry $d=\frac{|\Gamma_\text{1D,1}-\Gamma_\text{1D,2}|}{\Gamma_\text{1D,1}+\Gamma_\text{1D,2}}$ in waveguide decay rate between qubits is 0.14 (0.03) for type-I (type-II) from Table~\ref{tb:rabi-fitting}, and this affects the decay rate of probe qubit by at most $\sim2\%$.
\subsection{Fitting of Rabi oscillation curves}
The Rabi oscillation curves in Fig.~3a and Fig.~4d are modeled using a numerical master equation solver \cite{Johansson2012, Johansson2013}. The qubit parameters used for fitting the Rabi oscillation curves are summarized in Table~\ref{tb:rabi-fitting}. For all the qubits, $\GammaoneD$ was found from spectroscopy. In addition, we have done a time-domain population decay measurement on the probe qubit to find the total decay rate of $\Gammaone/2\pi = 1.1946$ MHz (95\% confidence interval $[1.1644, 1.2263]$ MHz, measured at 6.55 GHz). Using the value of $\GammaoneD/2\pi = 1.1881$ MHz (95\% confidence interval $[1.1550, 1.2211]$ MHz, measured at 6.6 GHz) from spectroscopy, we find the spurious population decay rate $\Gammaloss/2\pi =\Gammaone/2\pi -\GammaoneD/2\pi =  6.5\:$kHz (with uncertainty of 45.3 kHz) for the probe qubit. The value of spurious population decay rate is assumed to be identical for all the qubits in the experiment. 

The dephasing rate of the probe qubit is derived from time-domain population decay and Ramsey sequence measurements $\Gammaphi = \Gammatwo - \Gammaone/2$. In the case of the mirror qubits, the table shows effective single qubit parameters inferred from measurements of the dark state lifetime. We calculate single mirror qubit dephasing rates that theoretically yield the corresponding measured collective value. Assuming an uncorrelated Markovian dephasing for the mirror qubits forming the cavity we find $\Gamma_{\varphi,\text{m}} = \Gamma_{\varphi,\text{D}}$ (see App.~\ref{App:D}). Similarly, the waveguide decay rate of the mirror qubits is found from the spectroscopy of the bright collective state as $\Gamma_{\text{1D,m}} = \Gamma_{\text{1D,B}}/2$. The detuning between probe qubit and the atomic cavity ($\Delta$) is treated as the only free parameter in our model. The value of $\Delta$ sets the visibility and frequency of the Rabi oscillation, and is found from the the fitting algorithm.

\begin{table}[h]
\centering
\begin{threeparttable}[t]
  \centering
\begin{tabular}{M{2.5cm}M{2.3cm}M{1.5cm}M{1.5cm}M{1.5cm}M{1.5cm}M{1.5cm}} 
Type    & Qubits involved    & $\Gamma_\text{1D,p}/2\pi$ (MHz)    & $\Gamma_\text{1D,m}/2\pi$ (MHz) & $\Gamma_{\varphi,\text{p}}/2\pi$ (kHz)  & $\Gamma_{\varphi,\text{m}}/2\pi$ (kHz) & $\Delta/2\pi$ (MHz)\\ \hline \hline
I & Q$_2$, Q$_6$ & 1.19 & {13.4} & 191 & 210 & 1.0 \\
II  & Q$_1$, Q$_7$ & 0.87 &  96.7 & 332 & 581 & 5.9 \\  
Dark compound &  Q$_2$Q$_3$, Q$_5$Q$_6$     &   1.19 & 4.3 & 191 & 146 & 0.9 \\
Bright compound  &   Q$_2$Q$_3$, Q$_5$Q$_6$   & 1.19  & 20.2 & 191   &  253 & 1.4
 \end{tabular}
\end{threeparttable}
 \caption{\textbf{Parameters used for fitting Rabi oscillation curves} The first and second row are the data for 2-qubit dark states, the third and fourth row are the data for 4-qubit dark states made of compound mirrors.  Here, $\GammaoneDp$ ($\Gamma_\text{1D,m}$) is the waveguide decay rate and $\Gamma_{\varphi,\text{p}}$ ($\Gamma_{\varphi,\text{m}}$) is the pure dephasing rate of probe (mirror) qubit, $\Delta$ is the detuning between probe qubit and mirror qubits used for fitting the data.}
   \label{tb:rabi-fitting}
\end{table}

\section{Lifetime ($T_1$) and coherence time ($T_2^*$) of dark state}
\label{App:D}

\begin{table}[b]
\centering
\begin{threeparttable}[t]
  \centering
\begin{tabular}{M{2.5cm}M{2.3cm}M{2cm}M{2cm}} 
Type    & Qubits involved    & $\Gamma_{1,\text{D}}/2\pi$ (kHz)   & $\Gamma_{2,\text{D}}/2\pi$ (kHz)   \\ \hline \hline
I & Q$_2$, Q$_6$ & 210 & 366  \\
II  & Q$_1$, Q$_7$ & 581 & 838  \\  
Dark compound &  Q$_2$Q$_3$, Q$_5$Q$_6$     &   146    &    215  \\
Bright compound  &   Q$_2$Q$_3$, Q$_5$Q$_6$   &    253   &  376
 \end{tabular}
\end{threeparttable}
 \caption{\textbf{Decay rate and decoherence rate of dark states.} The first and second row are the data for 2-qubit dark states, the third and fourth row are the data for 4-qubit dark states made of compound mirrors.  Here, $\Gamma_{1,\text{D}}$ ($\Gamma_{2,\text{D}}$) is the decay (decoherence) rate of the dark state.}
   \label{tb:darkDynamics}
\end{table}
The dark state of mirror qubits belongs to the decoherence-free subspace in the system due to its collectively suppressed radiation to the waveguide channel. However, there exists non-ideal channels that each qubit is coupled to, and such channels contribute to the finite lifetime ($T_1$) and coherence time ($T_2^*$) of the dark state (See Table~\ref{tb:darkDynamics}). In the experiment, we have measured the decoherence rate $\Gamma_{2,\text{D}}$ of the dark state to be always larger than the decay rate $\Gamma_{1,\text{D}}$, which cannot be explained by simple Markovian model of two qubits subject to their own independent noise. We discuss possible scenarios that can give rise to this situation of $\Gamma_{2,\text{D}}>\Gamma_{1,\text{D}}$, with distinction of the Markovian and non-Markovian noise contributions.

There are two major channels that can affect the coherence of the dark state. First, coupling of a qubit to dissipative channels other than the waveguide can give rise to additional decay rate $\Gammaloss=\Gammaone - \GammaoneD$ (so-called non-radiative decay rate). 
This type of decoherence is uncorrelated between qubits and is well understood in terms of the Lindblad form of master equation, whose contribution to lifetime and coherence time of dark state is similar as in individual qubit case. 
Another type of contribution that severely affects the dark state coherence arises from fluctuations in qubit frequency, which manifest as pure dephasing rate $\Gammaphi$ in the individual qubit case. This can affect the decoherence of the dark state in two ways: (i) By accumulating a relative phase between different qubit states, this act as a channel to map the dark state into the bright state with short lifetime, and hence contributes to loss of population in the dark state; (ii) fluctuations in qubit frequency also induces the frequency jitter of the dark state and therefore contributes to the dephasing of dark state.

In the following, we model the aforementioned contributions to the decoherence of dark state. Let us consider two qubits separated by $\lambda/2$ along the waveguide on resonance, in the presence of fluctuations $\tilde\Delta_j(t)$ in the qubit frequency. 
The master equation can be written as $\dot{\hat{\rho}}=-i[\hat{H}/\hbar, \hat{\rho}] + \mathcal{L}[\hat{\rho}]$, where the Hamiltonian $\hat{H}$ and the Liouvillian $\mathcal{L}$ are given by
\begin{align}
    \hat{H}(t) &=\hbar \sum_{j = 1,2}{\tilde\Delta_j(t)}\sigmap^{(j)}\sigmam^{(j)}, \label{eq:DephasingHamiltonian} \\
    \mathcal{L}[\rho] &=\sum_{j,k = 1,2} \left[(-1)^{j-k}\Gamma_{\text{1D}} + \delta_{jk}\Gammaloss \right] \left(\sigmam^{(j)}\hat{\rho}\sigmap^{(k)} - \frac{1}{2}\{\sigmap^{(k)}\sigmam^{(j)}, \hat{\rho}\}\right).\label{eq:DephasingLiouvillian} 
\end{align}
Here, $\GammaoneD$ ($\Gammaloss$) is the decay rate of qubits into waveguide (spurious loss) channel. Note that we have assumed the magnitude of fluctuation $\tilde\Delta_j(t)$ in qubit frequency is small and neglected its effect on exchange interaction and correlated decay. We investigate two scenarios in the following subsections depending on the correlation of noise that gives rise to qubit frequency fluctuations.

\subsection{Markovian noise}

If the frequency fluctuations of the individual qubits satisfy the conditions for Born and Markov approximations, i.e. the noise is weakly coupled to the qubit and has short correlation time, the frequency jitter can be described in terms of the standard Lindblad form of dephasing \cite{Gardiner2004}. 

More generally, we also consider the correlation between frequency jitter of different qubits. Such contribution can arise when different qubits are coupled to a single fluctuating noise source. For instance, if two qubits in a system couple to a magnetic field $B_0+\tilde B(t)$ that is global to the chip with $ D_j \equiv \partial \tilde\Delta_j/\partial \tilde B$, the correlation between detuning of different qubits follows correlation of the fluctuations in magnetic field, giving
$\langle \tilde\Delta_1(t)\tilde\Delta_2(t+\tau) \rangle  = D_1D_2\langle \tilde B(t)\tilde B(t+\tau) \rangle\neq 0$.
The Liouvillian associated with dephasing can be written as \cite{Duan1998}
\begin{equation}
    \mathcal{L}_{\varphi,jk}[\hat{\rho}] = \frac{\Gamma_{\varphi,jk}}{2}\left(\sigmaz^{(j)}\hat{\rho}\sigmaz^{(k)} - \frac{1}{2}\left\{\sigmaz^{(k)}\sigmaz^{(j)}, \hat{\rho}\right\}\right),\label{eq:CorrDeph}
\end{equation}
where the dephasing rate  $\Gamma_{\varphi,jk}$ between qubit $j$ and qubit $k$ ($j = k$ denotes individual qubit dephasing and $j\neq k$ is the correlated dephasing) is given by
\begin{equation}
    \Gamma_{\varphi,jk} \equiv \frac{1}{2} \int_{-\infty}^{+\infty} \mathrm{d}\tau\:\langle \tilde\Delta_j(0)\tilde\Delta_k(\tau) \rangle.
\end{equation}
Here, the average $\langle\cdot\rangle$ is taken over an ensemble of fluctuators in the environment. 
Note that the correlated dephasing rate $\Gamma_{\varphi,jk}$ can be either positive or negative depending on the sign of noise correlation, while the individual pure dephasing rate $\Gamma_{\varphi,jj}$ is always positive.

After we incorporate the frequency jitter as the dephasing contributions to the Liouvillian, the master equation takes the form
\begin{equation}
    \dot{\hat{\rho}} = \sum_{j,k = 1,2}\left\{ \left[(-1)^{j-k}\Gamma_{\text{1D}} + \delta_{jk}\Gammaloss \right] \left(\sigmam^{(j)}\hat{\rho}\sigmap^{(k)} - \frac{1}{2}\{\sigmap^{(k)}\sigmam^{(j)}, \hat{\rho}\}\right)+\frac{\Gamma_{\varphi,jk}}{2}\left(\sigmaz^{(j)}\hat{\rho}\sigmaz^{(k)} - \frac{1}{2}\left\{\sigmaz^{(k)}\sigmaz^{(j)}, \hat{\rho}\right\}\right)\right\},\label{eq:master-full-dephasing}
\end{equation}
We diagonalize the correlated decay part of the Liouvillian describe the two-qubit system in terms of bright and dark states defined in Eq.~\eqref{eq:DarkBrightStates}. 
From now on, we assume the pure dephasing rate and the correlated dephasing rate are identical for the two qubits, and write $\Gammaphi\equiv \Gamma_{\varphi,11}=\Gamma_{\varphi,22}$, $\Gamma_{\varphi,\text{c}}\equiv \Gamma_{\varphi,12}=\Gamma_{\varphi,21}$. For qubits with a large Purcell factor ($\Gamma_\mathrm{1D} \gg \Gamma_{\varphi}, |\Gamma_{\varphi,\text{c}}|, \Gammaloss$), we can assume that the superradiant states $|\text{B}\rangle$ and $|\text{E}\rangle$ are only virtually populated \cite{Paulisch:2016ib} and neglect the density matrix elements associated with $|\text{B}\rangle$ and $|\text{E}\rangle$. Rewriting Eq.~\eqref{eq:master-full-dephasing} in the basis of $\{|\text{G}\rangle, |\text{B}\rangle, |\text{D}\rangle, |\text{E}\rangle\}$,
the dynamics related to dark state can be expressed as $\dot{\rho}_\mathrm{D,D} \approx -\Gamma_{1,\text{D}}\rho_\mathrm{D,D}$ and $\dot{\rho}_\mathrm{D,G} \approx -\Gamma_{2,\text{D}}\rho_\mathrm{D,G}$, where 
\begin{equation}
    \Gamma_{1,\text{D}}=\Gammaloss+\Gamma_{\varphi}-\Gamma_{\varphi,\text{c}},\quad \Gamma_{2,\text{D}}=\frac{\Gammaloss}{2}+\Gamma_{\varphi}.\label{eq:darkDynamics}
\end{equation}
Note that if the correlated dephasing rate $\Gamma_{\varphi,\text{c}}$ is zero, $\Gamma_{1,\text{D}}$ is always larger than $\Gamma_{2,\text{D}}$, which is in contradiction to our measurement result.

We estimate the decay rate into non-ideal channels to be $\Gammaloss/2\pi=6.5$\:kHz from the difference in $\Gammaone$ and $\GammaoneD$ of the probe qubit, and assume $\Gammaloss$ to be similar for all the qubits. 
Applying Eq.~\eqref{eq:darkDynamics} to measured values of $\Gamma_{2,\text{D}}$ listed in Table~\ref{tb:darkDynamics}, we expect that the pure dephasing of the individual qubit is the dominant decay and decoherence source for the dark state. In addition, we compare the decay rate $\Gamma_{1,\text{D}}$ and decoherence rate $\Gamma_{2,\text{D}}$ of dark states in the Markovian noise model and infer that the correlated dephasing rate $\Gamma_{\varphi,\text{c}}$ is positive and is around a third of the individual dephasing rate $\Gamma_{\varphi}$ for all types of mirror qubits.

\subsection{Non-Markovian noise}
In a realistic experimental setup, there also exists non-Markovian noise sources contributing to the dephasing of the qubits, e.g. $1/f$-noise or quasi-static noise \cite{Martinis2003, Ithier2005, Koch2007}. In such cases, the frequency jitter cannot be simply put into the Lindblad form as described above. In this subsection, we consider how the individual qubit dephasing induced by non-Markovian noise influences the decoherence of dark state. As shown below, we find that a non-Markovian noise source can lead to a shorter coherence time to lifetime ratio for the dark states, in a similar fashion to correlated dephasing. However, we find that the functional form of the visibility of Ramsey fringes is not necessarily an exponential for a non-Markovian noise source.

We start from the master equation introduced in Eqs.~\eqref{eq:DephasingHamiltonian}-\eqref{eq:DephasingLiouvillian} can be written in terms of the basis of $\{|\text{G}\rangle, |\text{B}\rangle, |\text{D}\rangle, |\text{E}\rangle\}$,
\begin{equation}
\label{eq:master2Q}
    \dot{\hat{\rho}} = -\frac{i}{\hbar}[\hat{H}, \hat{\rho}] + \left(2\GammaoneD + \Gammaloss\right) \mathcal{D}[\hat{S}_\text{B}]\hat{\rho} + \Gammaloss \mathcal{D}[\hat{S}_\text{D}]\hat{\rho},
\end{equation}
where the Hamiltonian is written using the common frequency jitter $\tilde\Delta_c(t)\equiv [\tilde\Delta_1(t)+\tilde\Delta_2(t)]/2$ and differential frequency jitter $\tilde\Delta_d(t)\equiv [\tilde\Delta_1(t)-\tilde\Delta_2(t)]/2$
\begin{equation}
    \hat{H}(t)/\hbar = \tilde\Delta_c(t)\left(2|\text{E}\rangle\langle \text{E}|+|\text{D}\rangle\langle \text{D}|+|\text{B}\rangle\langle \text{B}|\right)+\tilde\Delta_d(t)\left(|\text{B}\rangle\langle \text{D}|+|\text{D}\rangle\langle \text{B}|\right).\label{eq:Hamiltonian-commdiff}
\end{equation}
Here, $\hat{S}_\text{B}$ and $\hat{S}_\text{D}$ are defined in Eq.~\eqref{eq:BrightDarkSigmaSymm}.  
From the Hamiltonian in Eq.~\eqref{eq:Hamiltonian-commdiff}, we see that the common part of frequency fluctuation $\tilde\Delta_c(t)$ results in the frequency jitter of the dark state while the differential part of frequency fluctuation $\tilde\Delta_d(t)$ drives the transition between $|\text{D}\rangle$ and $|\text{B}\rangle$, which acts as a decay channel for the dark state. 

For uncorrelated low-frequency noise on the two qubits, the decoherence rate is approximately the standard deviation of the common frequency jitter $\sqrt{\langle \tilde\Delta_c(t)^2 \rangle}$. The decay rate in this model can be found by modeling the bright state as a cavity in the Purcell regime, and calculate the damping rate of the dark state using the Purcell factor as $\langle 4\tilde\Delta_d(t)^2/\Gamma_\mathrm{B}\rangle$. As evident, in this model the dark state's population decay rate is strongly suppressed by the large damping rate of bright state $\Gamma_\mathrm{B}$, while the dark state's coherence time can be sharply reduced due to dephasing.

\section{Shelving}
\label{App:E}

We consider the case of two identical mirror qubits of frequency $\omegaq$, separated by distance $\lambda/2$ along the waveguide. In addition to free evolution of qubits, we include a coherent probe signal from the waveguide in the analysis. In the absence of pure dephasing ($\Gammaphi=0$) and thermal occupancy ($\nth=0$), the master equation in the rotating frame of the probe signal takes the same form as Eq.~\eqref{eq:master2Q}, where the Hamiltonian containing the drive from the probe signal is written as
\begin{equation}
    \hat{H}/\hbar = \sum_{\mu = \textrm{B}, \textrm{D}}\left[-\domega\: \hat{S}_\mu^\dagger \hat{S}_\mu + \frac{\Omega_\mu}{2}\left( \hat{S}_\mu+ \hat{S}_\mu^{\dagger}\right)\right]
    ,\label{eq:2QB-halfwavelength}
\end{equation}
where $\hat{S}_\text{B}$ and $\hat{S}_\text{D}$ are defined in Eq.~\eqref{eq:BrightDarkSigmaSymm}, $\domega=\omegap-\omegaq$ is the detuning of the probe signal from the mirror qubit frequency,
$\Omega_\mu$ is the corresponding driven Rabi frequency.
Note that due to the symmetry of the excitations with respect to the waveguide, we see that $\Omega_\text{D} = 0$ and $\Omega_\text{B} = \sqrt{2}\Omega_1$, where $\Omega_1$ is the Rabi frequency of one of the mirror qubits from the probe signal along the waveguide.

Let us consider the limit where the Purcell factor $P_{\mathrm{1D}} = \Gamma_{\mathrm{1D}}/\Gamma'$ of qubits is much larger than unity (equivalent to $\Gamma_\mathrm{D} = \Gammaprime \ll \Gamma_\mathrm{B} = 2 \GammaoneD + \Gammaprime$) and the drive applied to the qubits is weak $\sub{\Omega}{B}\ll \sub{\Gamma}{B}$. Then, we can effectively remove some of the density matrix elements \footnote{
From the master equation, the time-evolution of part of the density matrix elements are approximately written as
\begin{align*}
\drhosub{E,E} &= -(\sub{\Gamma}{B}+\sub{\Gamma}{D})\rhosub{E,E} + \frac{i\Omega_\text{B}}{2}(\rhosub{B,E} - \rhosub{E,B}), \\
\drhosub{E,B} &= \left[i\delta\omega -\left( \sub{\Gamma}{B} + \frac{\sub{\Gamma}{D}}{2}\right)\right] \rhosub{E,B} + \frac{i\Omega_\text{B}}{2} (\rhosub{B,B} - \rhosub{E,E} + \rhosub{E,G}), \\
\drhosub{E,G} &= \left(2i\delta\omega -\frac{\sub{\Gamma}{B} + \sub{\Gamma}{D}}{2}\right) \rhosub{E,G} + \frac{i\Omega_\text{B}}{2} (\rhosub{B,G} + \rhosub{E,B}), \\
\drhosub{E,B} &= \drhosub{B,E}^*;\quad \drhosub{E,G} = \drhosub{G,E}^*.
\end{align*}
In the steady state, it can be shown that 
\begin{align*}
    \rhosub{E,E} &\sim \mathcal{O}(x^2) \rhosub{B,B} + \mathcal{O}(x^3) (\rhosub{B,G} - \rhosub{G,B})\\
    \rhosub{B,E} &\sim \mathcal{O}(x) \rhosub{B,B} + \mathcal{O}(x^2)\rhosub{G,B}\\
    \rhosub{G,E} &\sim \mathcal{O}(x^2) \rhosub{B,B} + \mathcal{O}(x)\rhosub{G,B}\\
\end{align*}
to leading order in $x\equiv \sub{\Omega}{B}/\sub{\Gamma}{B} < 1$, and hence we can neglect the contributions from $\rhosub{E,E}$, $\rhosub{B,E}$, $\rhosub{E,B}$, $\rhosub{G,E}$, $\rhosub{E,G}$ from the analysis in the weak driving limit. The probe power we have used in the experiment corresponds to $x\sim 0.15$, which makes this approximation valid.},
$$
\rhosub{E,E},\:\rhosub{B,E},\:\rhosub{E,B},\:\rhosub{G,E},\:\rhosub{E,G}\approx 0 
$$
and restrict the analysis to ones involved with three levels $\{|\text{G}\rangle, |\text{D}\rangle, |\text{B}\rangle\}$. In addition, the dark state $|\text{D}\rangle$ is effectively decoupled from $ |\text{G}\rangle$ and $|\text{B}\rangle$, acting as a metastable state. Therefore, we only consider the following set of the master equation:
\begin{align}
\drhosub{B,B} &\approx -\sub{\Gamma}{B}\:\rhosub{B,B} + \frac{i\Omega_\text{B}}{2}(\rhosub{B,G} - \rhosub{G,B})\label{eq:shelving-eq1}\\
\drhosub{B,G} &\approx \left(i\domega - \frac{\sub{\Gamma}{B}}{2}\right)\rhosub{B,G} + \frac{i\Omega_\text{B}}{2}(\rhosub{B,B} - \rhosub{G,G})\\
\drhosub{G,G} &\approx -\drhosub{B,B};\quad \drhosub{G,B}=\drhosub{B,G}^*\label{eq:shelving-eq2}
\end{align}
Using the normalization of total population $\rhosub{G,G} + \rhosub{D,D}+\rhosub{B,B}\approx 1$ with Eqs.~\eqref{eq:shelving-eq1}-\eqref{eq:shelving-eq2}, we obtain the approximate steady-state solution
\begin{equation}
     \langle \hat{S}_\mathrm{B} \rangle\approx\sub{\rho}{B,G}\approx- \frac{i \sub{\Omega}{B} (1-\sub{\rho}{D,D})}{\sub{\Gamma}{B}-2i\domega}.
\end{equation}
The input-output relation \cite{Lalumiere:2013io} is given as
\begin{align}
    \hat{a}_{\mathrm{out}} =  \hat{a}_{\mathrm{in}}  + \sqrt{\frac{\GammaoneD}{2}}\sigmam^{(1)} - \sqrt{\frac{\GammaoneD}{2}}\sigmam^{(2)} = \hat{a}_{\mathrm{in}} + \sqrt{\GammaoneD}\sub{\hat{S}}{B},
\end{align}
where $\ain$ is the input field operator and $\aout$ is the operator for output field propagating in the same direction as the input field (here, the input field is assumed to be incident from only one direction). The transmission amplitude is calculated as \begin{equation}
    t = \frac{\langle \hat{a}_\mathrm{out} \rangle}{\langle \hat{a}_\mathrm{in} \rangle} = 1 - \frac{(1-\rhosub{D,D})\GammaoneD}{-i\domega + \sub{\Gamma}{B}/2}
\end{equation}
where the relation $\Omega_\mathrm{1}/2 = -i\langle \hat{a}_{\mathrm{in}} \rangle  \sqrt{\Gamma_\mathrm{1D}/2}$ has been used.

In the measurement, we use the state transfer protocol to transfer part of the ground state population into the dark state. Following this, we drive the $|\text{G} \rangle \leftrightarrow |\mathrm{B}\rangle$ transition by sending a weak coherent pulse with a duration 260 ns into the waveguide, and recording the transmission spectrum. As a comparison, we also measure the transmission spectrum when the mirror qubits are in the ground state, which corresponds to having $\rho_\mathrm{D,D} = 0$. The transmittance in the two cases (Figure~3d) are fitted with identical parameters for $\Gamma_\mathrm{1D}$ and $\Gamma_\mathrm{B}$. The dark state population $\rho_\mathrm{D,D}$ following the iSWAP gate is extracted from the data as 0.58, which is lower than the value (0.68) found from the Rabi oscillation peaks (Figure~3a). The lower value of the dark state population can be understood considering the finite lifetime of dark state ($757\:\mathrm{ns}$), which leads to a partial population decay during the measurement time (the single-shot measurement time is set by the duration of the input pulse). It should be noted that the input pulse has a transform-limited bandwidth of $\sim 3.8$\:MHz, which results in frequency averaging of the spectral response over this bandwidth. For this reason, the on-resonance transmission extinction measured in the pulsed scheme is lower than the value found from continuous wave (CW) measurement (Fig.~\ref{fig:schematic}c).

\end{document}